\documentclass{mn2e}
\usepackage{psfig}
\usepackage{mnras_cite}
\newcommand{\redshifthi}{\left( \frac{1+z}{20} \right)}   
\newcommand{\redshiftnine}{\left( \frac{1+z}{10} \right)}

\newcommand{\htwo}{{\rm H_{2}}}

\newcommand{\be}{\begin{equation}}
\newcommand{\ba}{\begin{eqnarray}}
\newcommand{\ee}{\end{equation}}
\newcommand{\ea}{\end{eqnarray}}

\def\gsim{\;\rlap{\lower 2.5pt
 \hbox{$\sim$}}\raise 1.5pt\hbox{$>$}\;}
\def\lsim{\;\rlap{\lower 2.5pt
   \hbox{$\sim$}}\raise 1.5pt\hbox{$<$}\;}

\begin{document}

\title[Fossil HII Regions: Self-Limiting Star Formation at High Redshift]{Fossil HII Regions: Self-Limiting Star Formation at High Redshift} 
\author[Oh \& Haiman]{S. Peng Oh$^{1}$ \& Zolt\'an Haiman$^{2}$ \\
$^1$Theoretical Astrophysics, Mail Code 130-33, Caltech, Pasadena, 
CA 91125, USA.\\
$^2$Department of Astronomy, Columbia University, 550 West 120th Street, New York, NY 10027, USA.}

\maketitle

\begin{abstract}
Recent results by the {\it WMAP} satellite suggest that the
intergalactic medium was significantly reionized at redshifts as high
as $z\sim 17$.  At this early epoch, the first ionizing sources likely
appeared in the shallow potential wells of mini--halos with virial
temperatures $T_{\rm vir} < 10^{4}$K. Once such an ionizing source
turns off, its surrounding HII region Compton cools and
recombines. Nonetheless, we show that the ``fossil'' HII regions left
behind remain at high adiabats, prohibiting gas accretion and cooling
in subsequent generations of mini--halos. Thus, early star formation
is self--limiting. 
We quantify this effect to show that star formation in mini--halos
cannot account for the bulk of the electron scattering opacity measured by {\it WMAP}, which must be due to more massive objects. We
argue that gas entropy, rather than IGM metallicity, regulates the
evolution of the global ionizing emissivity, and impedes full
reionization until lower redshifts. We discuss several important
consequences of this early entropy floor for reionization. It reduces
gas clumping, curtailing the required photon budget for reionization. An entropy floor
also prevents $\htwo$ formation and cooling, due to reduced gas
densities: it greatly enhances feedback from UV photodissociation of
$\htwo$. An early X-ray background would also furnish an entropy floor
to the entire IGM; thus, X-rays impede rather than enhance $\htwo$
formation. Future 21cm observations may probe the topology of fossil
HII regions.
\end{abstract}

\section{Introduction}

The high optical depth $\tau = 0.17 \pm 0.04$ detected by the
Wilkinson Microwave Anisotropy Probe ({\it WMAP}) satellite has lent
greater credence to the notion of an early period of star formation
and reionization, $z_{r}= 17 \pm 8$ \cite{kogut,spergel}. If indeed
the first stars formed at high redshift $z \sim 20$, they are expected
to form in mini--halos\footnote{For the purposes of this paper, a
mini--halo is defined as any halo with $T_{\rm vir} < 10^{4}$K which
cannot cool via atomic line cooling.} with shallow potential wells, in
which ${\rm H_{2}}$ cooling is dominant \cite{abel,bromm}. More
massive halos with $T_{\rm vir} > 10^{4}$K, in which collisional
ionization and line cooling can operate, are expected to be very rare
at these redshifts. Modulo the effects of UV feedback on $\htwo$
formation and cooling, stars forming in such halos could therefore
play a dominant role in an early reionization epoch. A great deal of
effort has gone into assessing the impact of UV feedback on $\htwo$
cooling, as well as the counter-vailing effects of positive feedback
effects such as an early X-ray background (e.g.,
\scite{haimanh2,ciardi,ricotti,machacek}).

The main point of this paper is that the population of mini--halos is
likely to be considerably sparser than previously assumed. This is
because mini--halo formation is strongly suppressed even inside the
fossil HII regions of dead ionizing sources. Although such HII regions
recombine and cool by Compton scattering with cosmic microwave
background (CMB) photons, they cannot cool to back to the temperature
of the undisturbed intergalactic medium (IGM). Strong Jeans mass
filtering takes place (Gnedin 2000), and subsequent mini--halos will
no longer be able to accrete gas due to the smoothing effects of
finite gas pressure. {\it Thus, once any patch of the universe is
ionized, it can no longer host any more mini--halos, even if it
subsequently cools and recombines}. In effect, the birth of the first
stars leads to the demise of the mini--halo population: only one
generation of stars can form within these shallow potential wells.

While many authors have noted and commented on the Jeans mass
filtering after full reionization \cite{bullock,benson,somerville},
the Jeans filtering in fossil HII regions after ionizing sources have
turned off has not been studied. We argue that it will strongly
suppress the mini--halo population, with the following interesting
consequences:

\begin{itemize}
{\item \bf Impact of Mini--Halo Population on Reionization.} Since the
recombination time is shorter than the Hubble time at high redshift,
$t_{rec} \ll t_{H}$, and the ionizing sources, expected to be massive
stars, have short lifetimes, $t_{MS} \sim 3 \times 10^{6} {\rm yr} \ll
t_{H}$, many generations of star formation are required to keep a
given patch of the IGM ionized. However, once the first generation of
stars born in mini--halos dies out, subsequent generations will not be
able to form in mini--halos in a previously reionized patch of IGM,
even after the patch cools and recombines.  This effect inevitably
produces a non--monotonic reionization history with an early peak of
partial reionization, followed by recombination and eventual full
reionization.  This is qualitatively similar to the reionization
history derived by Cen (2003) and Wyithe \& Loeb (2003), although the
physical reason for the non--monotonic evolution in the present work
is different (entropy injection, rather than a pop III to pop II
transition caused by a universal metallicity increase).  Regardless of
the extent of UV feedback effects on $\htwo$ production and cooling,
the majority of stars which reionized the universe were hosted by more
massive halos able to survive Jeans mass filtering, $T_{vir} > {\rm
few} \times 10^{4}$K.
 
{\item \bf Photon Budget for Reionization.} Dense gas that collects in
mini--halos can form a considerable sink of ionizing photons by
boosting the overall effective gas clumping factor $C_{II}= \langle
n_{e}^{2} \rangle / \langle n_{e} \rangle^{2}$
\cite{haimanabelmadau,barkanaloeb,shapiro03}. The clumping factor
$C_{II}$ increases rapidly at low redshift as structure formation
progresses; thus, despite the higher mean gas density $n \propto
(1+z)^{3}$ at high redshift, the recombination time $t_{rec}=1/(\alpha
n_{e} C_{II})$ does not evolve strongly from $z=10-20$. However, if
the universe is filled up with fossil HII regions at high redshift,
the photon budgets required to subsequently ionize it and to keep it
ionized are much lower, since gas clumping in mini--halos is strongly
suppressed.

{\item \bf Source Clustering.} Since preheating boosts the threshold
mass required for efficient gas cooling and star formation, it
increases the mean bias of the early proto-galaxy population by a
factor of a few (e.g., see figure 2 of \scite{ohetal03}).  This is
likely to increase the clustering amplitude of background fluctuations
due to these faint unresolved early proto-galaxies, such as: the
free-free background due to ionized halos \cite{oh99,oh_mack},
Sunyaev-Zeldovich fluctuations due to high redshift HII regions and
supernovae winds \cite{knoxetal99,ohetal03,santos}, and the IR
background due to stellar emission
\cite{santos_mike,ferrara,haiman_midIR}. The increase in clustering
bias boosts the amplitude of rms fluctuations by a factor of
several. This may be necessary, for instance, if thermal SZ
fluctuations from high-redshift supernovae are to account for the
small-scale CMB anisotropies observed by BIMA and CBI \cite{ohetal03}.
 
{\item \bf Star Formation History/Metallicity/IMF.} Initially, each
mini--halo is expected to harbor a single massive, metal--free star
(no fragmentation is seen in numerical simulations by Abel et al. 2000
and Bromm et al. 2002). Because of the preheating, no further
mini--halos can form in the entire comoving volume reionized by this
mini--halo.  As a result, the first generation of stars is expected to
form as a population of single, isolated stars.  Star formation ensues
again only once more massive halos with deeper potential wells
aggregate. These rare, high density peaks are likely to coincide with
the highly biased regions where the first isolated stars had already
formed. At the time of the formation of deeper potential wells, such
sites are inevitably already polluted with metals, unless the first
stars collapsed directly to black holes without associated metal
production. As a result, the transition from pop III to pop II (metal
free to normal) stellar populations is associated with halo mass
scale.  This is in contrast to scenarios (Cen 2003; Wyithe \& Loeb
2003) where star formation can continually proceed in lower density
peaks, which are far from the initial sites of star formation and
still contain relatively pristine gas (allowing metal-free star
formation to continue to relatively low redshifts). Given the factor
of $\sim 10-20$ difference in the ionizing photon production
efficiency per unit mass between Pop III and Pop II stars (Tumlinson
\& Shull 2000; Bromm, Kudritzki \& Loeb 2001; Schaerer 2002), this
will have important consequences for the redshift evolution and
topology of reionization.

{\item \bf 21cm Observations.} It has been suggested that mini--halos
will be observable at the redshifted 21cm line frequency both in
emission \cite{mh1,mh2}, and absorption \cite{furlanetto}. If
mini--halo formation is strongly suppressed, these two windows on
small-scale structure during the cosmological Dark Ages will
disappear. 21cm observations of mini--halos therefore provide an
interesting probe of the topology of fossil HII regions: if
mini--halos are seen, that comoving patch of the IGM has {\it never}
been ionized. This is complementary to other observations such as
Gunn-Peterson absorption, which only probe the instantaneous
ionization state.

\end{itemize}

The rest of this paper is organized as follows. In
\S\ref{section:entropy_floor}, we compute the ``entropy floor'' due to
early mini--halos. In \S\ref{section:halo_profiles}, we calculate the
effects of this entropy floor on the gas density profile in
mini--halos. In \S\ref{section:global}, we discuss and quantify
several important effects of this entropy floor for reionization, and
in \S\ref{sec:reionhist} we explicitly calculate the effect on the
reionization history. We summarize our findings and discuss their
implications in \S\ref{section:conclusions}.  Throughout this paper,
we adopt the background cosmological parameters as measured by the
{\it WMAP} experiment (Spergel et al. 2003, Tables 1 and 2),
$\Omega_m=0.29$, $\Omega_{\Lambda}=0.71$, $\Omega_b=0.047$, $h=0.72$
and an initial matter power spectrum $P(k) \propto k^n$ with $n=0.99$
and normalization $\sigma_8=0.9$.

\section{Entropy Floor due to Early Reionization}
\label{section:entropy_floor}

\subsection{Entropy floor in Fossil HII regions} 

Early reionization introduces an 'entropy floor' in the intergalactic
medium which impedes gas accretion and cooling in halos with $T_{\rm
vir} < 10^{4}$K which cannot excite atomic cooling. Early reionization
even introduces Jeans smoothing effects in halos with $T_{vir} >
10^{4}$K, since the strength of the accretion shock is weaker in
preheated gas, and a deeper potential well is required to
collisionally ionize the gas. The overall effect is to introduce a
substantial core in the gas density profile. The effect is similar to
the presumed preheating of the IGM by galactic winds or AGN outflows
at low redshift. There, the finite entropy of the gas introduces a
core in group and cluster gas density profiles and is thought to be
responsible for the deviation from self-similarity in the observed
cluster $L_{X}-T_{X}$ relation (for a recent review, see
\scite{rosatietal2002}). Entropy is more fundamental than gas
temperature in low-redshift clusters and high-redshift mini--halos: in
both cases, entropy is conserved, since $t_{\rm cool} \gg t_{H}$. We
exploit this analogy, and deliberately employ language and techniques
from the well-studied low redshift case to study high-redshift
mini--halos. Here, we calculate the expected level of the 'entropy
floor' due to early reionization.

\begin{figure}
\psfig{file=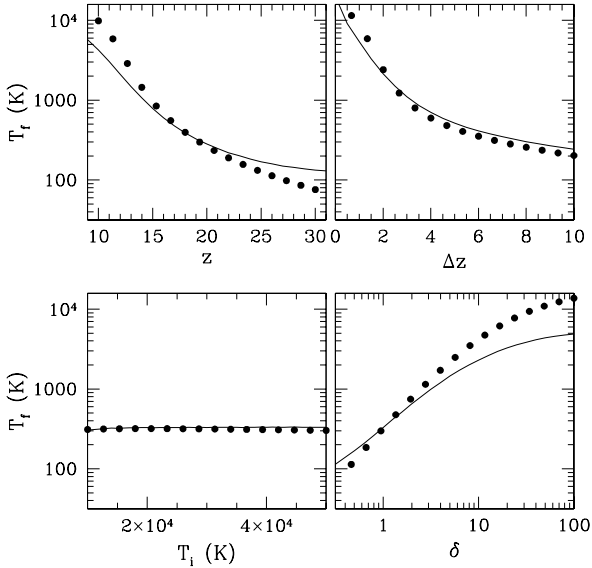,width=80mm}
\caption{The parameter dependence of the final temperature of a parcel
of isochorically cooling gas. The solid lines indicate the results
from the full non-equilibrium chemistry code, while the points
indicate the approximate analytic solution from equations
\ref{eqn:T_analytic} and \ref{eqn:t_rec}, which is in excellent
agreement. Our fiducial patch has an initial temperature
$T_{i}=20,000$K, cools at $z=19$, for a time interval $t=0.6
t_{H}(z=19)=1.9 \times 10^{8}$yr (corresponding to roughly $\delta z
\approx 7$), and lies at an overdensity $\delta=1$.  The gas cools
most effectively at high redshift (when the Compton cooling time is
short) and at low overdensity (when it decouples from the CMB at late
times). For z=19 and $\delta=10$, (which is perhaps characteristic of
gas being accreted onto halos) the gas remains at $\sim {\rm few}
\times 10^{3}$K.}
\label{fig:temp_params}
\end{figure}

Reionization is likely to be a highly stochastic process where sources
ionize a patch of the IGM and then fade; the fossil HII region
subsequently cools and recombines. What is the final temperature a
fossil HII region can cool to? At the high redshifts of interest,
Compton cooling off the CMB is by far the dominant source of gas
cooling in the low density IGM (e.g., at $z=19$ and $T=10^{4}$K,
$t_{C} \sim 0.1 t_{cool} \delta$, where $\delta$ is the gas
overdensity, and $t_{cool}$ is the atomic cooling timescale).  The
final gas temperature is therefore determined by the competition
between Compton cooling and hydrogen recombination: after the gas
recombines, it 'decouples' from the CMB, and Compton cooling is no
longer efficient. The Compton cooling timescale is independent of
density and temperature:
\begin{equation}
     t_{C}=3 m_{e}c (4 x_{e} \sigma_{T} a T_{\rm CMB}^{4})^{-1}= 1.4
     \times 10^{7} \redshifthi^{-4} x_{e}^{-1} {\rm yr},
\end{equation}
whereas the recombination time 
\begin{equation}
t_{\rm rec} = 3.9 \times 10^{7} \redshifthi^{-3} \delta^{-1} x_{e}^{-1} {\rm yr}
\end{equation}
is shorter for higher density gas. Thus, denser gas 'decouples' from
the CMB earlier and freezes out at {\it higher} temperatures. For the
redshifts of interest, $t_{\rm rec} (\delta=1) > t_{C}$, allowing for
substantial cooling below $\sim 10^{4}$K.

We can study the temperature evolution of a cooling and recombining
parcel of gas $T(t,T_{i},\delta,z)$ as a function of time $t$ (or
redshift interval $\delta z$), initial temperature $T_{i}$, gas
overdensity $\delta_{i}$ and redshift $z$. In Figure 2 we show the
dependence of the final temperature on these parameters by solving the
full set of chemical evolution equations for primordial gas. We
neglect $\htwo$ chemistry as a trace UV flux is sufficient to
photodissociate $\htwo$ at these low IGM densities. Our fiducial patch
has an initial temperature $T_{i}=20,000$K, cools at $z=19$, for a
time interval $t=0.6 t_{H}(z=19)$ (corresponding to roughly $\delta z
\approx 7$), and lies at an overdensity $\delta=1$; each panel shows
the dependence on one of the variables holding the others fixed. We do
not take into account time-dependent variation of the gas density
arising from infall onto non-linear structures or Hubble expansion,
which would provide adiabatic heating/cooling of the gas; however the
effects of this may be roughly estimated by considering some mean
overdensity $\delta$ of the gas during its evolution. The final
temperature is almost independent of $T_{i}$, as long as $t \gg t_{C}$
and the gas was initially fully ionized. However, the final
temperature declines at higher redshifts and low overdensities
$\delta$, since $t_{C}/t_{\rm rec} \propto (1+z)^{-1} \delta$, and the
gas cools faster than it recombines.

In Appendix I, we develop an analytic expression for
$T(t,T{i},\delta,z)$ taking into account only Compton cooling and
recombination which is remarkably accurate; it is given by the points
in Figure \ref{fig:temp_params}. This allows one to estimate the final
temperature quickly without evolving the coupled differential
equations. It deviates from the true answer significantly only at high
overdensity and/or low redshift, when atomic line cooling becomes
comparable to Compton cooling. In any case, for metal-free gas, line
cooling alone can only cool the gas down to $\sim 5000$K. The only
processes which could alter the final entropy of the gas significantly
are $\htwo$ cooling and metal line cooling. These are more efficient
than Compton cooling only in regions of high overdensity, in the halo
core, and so do not affect the initial entropy floor.

\begin{figure}
\psfig{file=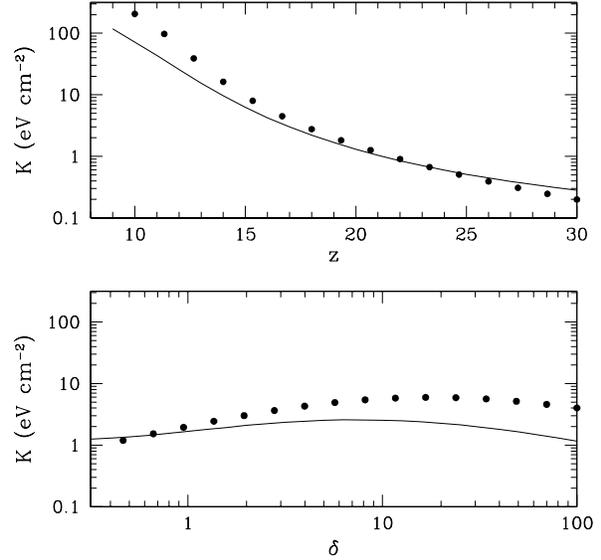,width=80mm}
\caption{Dependence of the final entropy $K=T/n^{2/3}$ on redshift $z$
and overdensity $\delta$. Points correspond to the analytic solution
from equations \ref{eqn:T_analytic} and \ref{eqn:t_rec}. As in Figure
\ref{fig:temp_params}, our fiducial patch has
$T_{i}=1,\delta=1,z=19,t=0.5 t_{H}(z=19)$. The dependence on $T_{i}$
and $t$ are the same as in Figure \ref{fig:temp_params}. The gas
retains more entropy at lower redshift, since Compton cooling is less
efficient. However, the final entropy shows only a weak dependence on
overdensity $\delta$: the gas cools out at higher temperatures at
higher densities, and the increased temperature and density roughly
cancel.}
\label{fig:entropy_params}
\end{figure}

Once cooling becomes ineffective, the gas evolves adiabatically. In
accordance with convention, we shall refer to the quantity:
\begin{equation}
K= \frac{T}{n^{2/3}} = 100 \left( \frac{k_{B} T}{1 {\rm eV}} \right)
\left(\frac{n_{e}}{10^{-3} {\rm cm^{-3}}} \right)^{-2/3} {\rm eV\,
cm^{2}}
\end{equation}
as the 'entropy' of the gas, even though it is not strictly the
thermodynamic entropy $S \propto {\rm log(K)}$. This is a useful
convention because $K$ is conserved when the gas evolves
adiabatically. Thus, it is conserved during Hubble expansion, as well
as during accretion onto halos (provided the accretion shock is weak,
gas will be accreted isentropically). In Figure
\ref{fig:entropy_params}, we show the dependence of the final entropy
$K$ on redshift (the dependence on $T_{i},t$ follow simply from Figure
\ref{fig:temp_params}). The gas has significantly lower entropy at
higher redshift, since Compton cooling is more efficient.  However,
the final entropy at a given redshift depends only weakly on the
overdensity $\delta$. Denser gas remains at higher temperature, since
it recombines faster than it cools, and the combination of higher
temperature and higher density roughly cancel $K \propto T
\delta^{-2/3}$. We can therefore ignore the weak $\delta$ dependence
of $K$, and assume that it depends only on $z$.  {\it The entropy
floor is thus roughly independent of the details of structure
formation and gas density distribution.}

\begin{figure}
\psfig{file=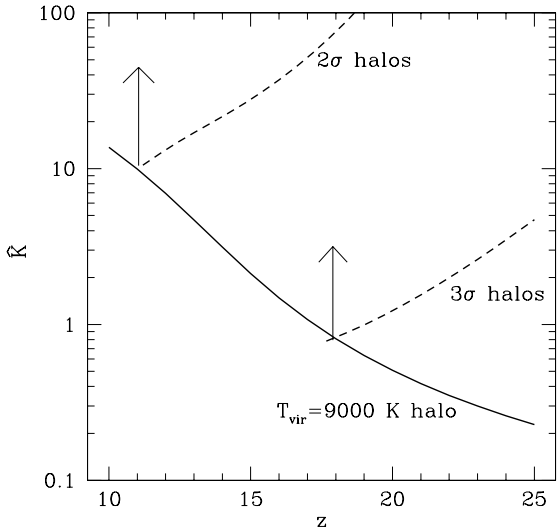,width=80mm}
\caption{The evolution of the dimensionless entropy parameter $\hat{K}
\equiv K_{\rm IGM}/K_{o}$ as a function of redshift, assuming the gas
cools at the mean IGM density $\delta=1$. Here $K_{o}$ is the gas
entropy at the virial radius due to shock heating alone, and $K_{\rm
IGM}$ is the entropy of the fossil HII regions as computed in Figure
\ref{fig:entropy_params}. The solid line describes a $T_{\rm
vir}=9000$K halo, the most massive in which atomic cooling is still
not important. It therefore defines a lower limit on $\hat{K}$. The
entropy parameter can be simply scaled ($\hat{K} \propto T_{\rm
vir}^{-1}$) for halos with lower virial temperatures. The dashed lines
show $\hat{K}$ for 2$\sigma$ and 3$\sigma$ halos at that redshift; the
vast majority of mini--halos have $\hat{K} >1$.}
\label{fig:Khat_z}
\end{figure}

It is useful to define a parameter which compares the IGM entropy
floor to the entropy generated by gravitational shock heating
alone. Let us define the quantity $\hat{K} \equiv K_{\rm IGM}/K_{o}$,
where $K_{o} = T_{\rm vir}/n(r_{\rm vir})^{2/3}$, and $n(r_{\rm
vir})=(\Omega_{b}/\Omega_{m})\rho_{NFW}(r_{\rm vir})/(\mu m_{p})$ (and
$\rho_{\rm NFW}$ is the NFW \cite{NFW} dark matter density
profile). This is the entropy due to shock heating alone at the virial
radius; the justification will become clearer in the ensuing
section. As $\hat{K}$ increases, the Jeans smoothing effects due to
finite IGM entropy become increasingly more significant. This
parameter $\hat{K}(K_{IGM},T_{\rm vir},z)$ will be used extensively in
the following sections, and it is useful here to get a sense of what
values of $\hat{K}$ are expected. In Figure \ref{fig:Khat_z}, we plot
$\hat{K}$ for a $T_{\rm vir}=9000$K halo as a function of redshift,
using $K_{\rm IGM}(z)$ shown in Figure \ref{fig:entropy_params}. Since
$\hat{K} \propto T_{\rm vir}^{-1}$, and a $T_{\rm vir}=9000$K halo is
about the most massive that would still not experience atomic cooling,
the solid line depicts a lower limit on $\hat{K}$ for all
mini--halos. Appropriate values for $\hat{K}$ for smaller halos can be
read off simply from the $\hat{K} \propto T_{\rm vir}^{-1}$
scaling. We see that for most halos, $\hat{K} > 1$; although
$K_{IGM}(z)$ falls at high redshift, this is somewhat offset by the
fact that typical potential wells are much shallower, and thus $K_{o}$
also falls rapidly (as seen by the dashed lines, which depict
$\hat{K}$ for 2$\sigma$ and 3$\sigma$ fluctuations).

\begin{figure}
\psfig{file=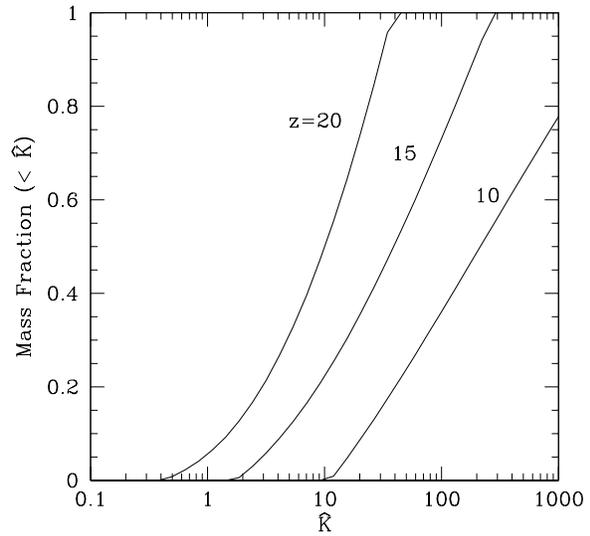,width=80mm}

\caption{The mass weighted fraction of mini--halos which have entropy
parameters $< \hat{K}$, for $z=10,15,20$, assuming $K_{\rm IGM}(z)$
shown in Figure \ref{fig:entropy_params}. Virtually all halos have
$\hat{K} > 1$, and the median value of $\hat{K}$ is substantially
higher. Thus, the vast majority of halos accrete gas isentropically.}
\label{fig:mass_frac_Khat}
\end{figure}

Typical values for $\hat{K}$ are illustrated further in Figure
\ref{fig:mass_frac_Khat}, where we show the mass weighted fraction of
mini--halos which have entropy parameters less than a given $\hat{K}$,
given by:
\begin{equation}
f(< \hat{K},z)= \frac{\int^{M(\hat{K})}_{M_{J}} dM M (dn/dM)} 
{\int^{M_{u}}_{M_{J}} dM M (dn/dM)}
\end{equation} 
where $M(\hat{K})$ is the mass corresponding to $\hat{K}$ for a given
$K_{\rm IGM}(z)$, $M_{J}(z)$ is the cosmological Jeans mass (e.g., see
\scite{barkana_loeb_review}), $M_{u}(z)$ is the mass corresponding to
a $T_{\rm vir}=10^{4}$K halo, and $dn/dM$ is the Press-Schechter mass
function. Virtually all halos have $\hat{K} > 1$ in fossil HII regions
at all redshifts of interest ($z\lsim 20$), and median values of
$\hat{K}$ are much higher. We shall soon see that such halos are
subject to very substantial Jeans smoothing effects.

\subsection{Entropy Floor due to X-rays}
\label{section:X_ray_floor}

Reionization by X-rays \cite{oh2001,venkatesan} would produce a warm
(few $\times 100-1000$K), weakly ionized IGM with an entropy floor
similar to that in fossil HII regions. Such X-rays could arise from
supernovae, AGN, or X-ray binaries, or in more exotic models with
decaying massive sterile neutrinos (Hansen \& Haiman 2003). The
universe is optically thick to all photons with energies:
\begin{equation}
E < E_{\rm thick} = 1.8 \left( \frac{1+z}{15} \right)^{0.5} x_{\rm
HI}^{1/3} {\rm keV}
\label{eqn:Ethick}
\end{equation}
where $x_{\rm HI}$ is the mean neutral fraction, and we have assumed
$\sigma_{\nu} \propto \nu^{-3}$. Thus, {\it all} energy radiated below
$E_{\rm thick}$ will be absorbed. The relative efficiency of UV
photons and X-rays in setting an entropy floor deserves detailed
separate study; here we discuss some salient points. UV photons are an
'energetically extravagant' means of producing an entropy floor. Most
of the energy injected by UV photons is lost to recombinations and
Compton cooling at high redshift; we see in Figure
\ref{fig:temp_params} that the gas typically Compton cools to $T_{\rm
floor} \sim {\rm few} \times 100$K at the redshifts of interest. Thus,
only $\sim 10^{-2}$ of the heat injected by UV photons is eventually
utilized in setting the entropy floor; all energy expended in heating
the gas above $T_{\rm floor}$ is 'wasted' (it is dumped into the CMB,
where it may eventually be observable as a spectral distortion, Fixsen
\& Mather 2002). On the other hand, in the weakly ionized gas produced
by X-rays, both the recombination time and the Compton cooling time
are longer than the Hubble time. In particular,
\begin{equation}
t_{C} = 1. \left(\frac{x_{e}}{0.1} \right)^{-1}
\left(\frac{1+z}{15}\right)^{-2.5} t_{H}
\end{equation}
and almost none of the entropy injected is lost to
cooling. Furthermore, a larger fraction of the X-ray energy goes
toward heating rather than ionization (in general, few$\times 0.1$ of
the energy of the hot photo-electron created by an X-ray goes toward
heating; this fraction quickly rises toward unity as the medium
becomes progressively more ionized \cite{shull}). Thus, if
$\epsilon_{\rm bol}({\rm X-ray})/\epsilon_{\rm bol}({\rm UV}) \ge
10^{-2}$ (where $\epsilon_{\rm bol}$ is the comoving emissivity),
X-rays could be comparable or even more effective than UV photons in
setting an entropy floor.  The relative emissivities of UV and X-rays
is unknown, but for supernovae could be as high as \cite{oh2001}:
\begin{equation}
\frac{\epsilon_{\rm bol}({\rm X-ray})}{\epsilon_{\rm bol}({\rm UV})} 
\approx 0.1 \left(\frac{f_{\rm esc}}{0.1} \right)^{-1} \left( \frac{f_{X}}{0.1}\right) 
\end{equation} 
where $f_{\rm esc}$ is the escape fraction of ionizing UV photons from
the host halo, and $f_{X}$ is the fraction of supernova explosion
energy which goes into soft X-rays, either through inverse Compton
scattering of CMB photons by relativistic electrons \cite{oh2001}, or
free-free emission from the hot SN remnant.

To make a quick estimate, let us (fairly conservatively) assume that
$f_{X} \sim 3\%$ of the explosion energy of a supernova goes into soft
X-rays. Of this, $f_{\rm heat} \sim 50\%$ of the energy goes into
heating; the rest goes into secondary ionizations and atomic
excitations. A supernova releases $E_{Z} \sim 0.5$MeV in explosion
energy per metal baryon, relatively independent of metallicity (a Pop
III 'hypernova' produces $\sim 100$ times more energy, but also $\sim
100$ times more metals than a standard type II SN). X-ray heating thus
results in an entropy floor $K_{IGM} \approx (f_{X} E_{Z} f_{\rm heat}
\bar{Z})/{n^{2/3}}$, or:
\begin{equation}
K_{IGM} \approx 20 \, {\rm eV cm^{2}}  \left( \frac{f_{X}}{0.03}
\right) \left( \frac{f_{\rm heat}}{0.5} \right) \left(
\frac{\bar{Z}}{10^{-3} {\rm Z_{\odot}}} \right)
\left(\frac{1+z}{15}\right)^{-2} 
\label{eqn:KIGM_Xray}
\end{equation}
where $\bar{Z}$ is the mean metallicity of the universe. This is
comparable to the entropy of fossil HII regions, but has a much larger
filling factor, of order unity. A metallicity of $\bar{Z} \sim 10^{-3}
{\rm Z_{\odot}}$ corresponds to roughly $\sim 1$ ionizing photon per
baryon in the universe, both for Pop II and Pop III stars. Due to
recombinations, the filling factor of ionized regions will be of
course considerably less than unity. However, the filling factor of
{\it fossil} HII regions, which is the relevant quantity, will also be
less than unity by a factor $f_{\rm overlap}$, which represents the
overlap of HII regions with fossil HII regions which have
recombined. Energy spent in reionizing fossil HII regions is 'wasted'
in terms of establishing an entropy floor. This factor is likely to be
large because early galaxy formation is highly biased, and higher mass
halos (which can resist feedback effects) will be born in regions
already pre-reionized by earlier generations of mini--halos. A fossil
HII filling factor of order unity with little overlap can only be
achieved if the formation of mini--halos is highly synchronized (see
discussion in section \ref{sec:reionhist}).

The mean free path of X-ray photons generally exceeds the mean
separation between sources, becoming comparable to the Hubble length
for $\sim 2$keV photons. The entire universe is thus exposed to a
fairly uniform X-ray background, and the entire IGM acquires a uniform
entropy floor, with an amplitude which scales with the amount of star
formation, as in equation (\ref{eqn:KIGM_Xray}). Even regions far from
sites of star formations, which have never been engulfed in an HII
region, will be affected. By contrast, in pure UV reionization
scenarios, there are large spatial fluctuations in entropy, which
depend on the topology of reionization, and the redshift at which a
comoving patch was last ionized.

Because of the relative uncertainty of the amplitude of the X-ray
background, we use the entropy floor associated with fossil HII
regions in the rest of this paper. The mass fraction of affected
mini--halos scales with the filling factor of fossil HII regions. This
is therefore a minimal estimate; the filling factor could approach
unity if X-rays are important.

We now consider the effects of a finite entropy floor on mini--halo
gas density profiles.

\section{Gas Density Profiles}
\label{section:halo_profiles}

Once Compton cooling and radiative cooling become inefficient, the gas
evolves adiabatically. We can therefore compute static equilibrium
density profiles, and see how they change as a function of the entropy
floor. The models we construct are in the spirit of
\scite{voitetal,tozzinorman,oh_benson,babuletal}, which match
observations of low-redshift cluster X-ray profiles well.

Naively, one might assume that only gas which remains at temperatures
comparable to the virial temperatures of mini--halos would suffer
appreciable Jeans smoothing. Since in many cases the IGM can cool down
to ${\rm few} \times 100$K, one might assume that this level of
preheating would have negligible effects on the density profile of gas
in mini--halos. This is false: the important quantity is not the
temperature but the entropy of the gas. Since gas in the IGM is heated
at low density, it has comparatively high entropy. Gas at mean density
which is heated to temperatures:
\begin{equation}
T_{\rm IGM} > 90 \left( \frac{T_{\rm vir}}{3000 {\rm K}} \right)
\left( \frac{\delta}{200} \right)^{-2/3} {\rm K} 
\end{equation}
(where $\delta$ is the overdensity of the gas in the mini--halo in the
absence of preheating) will have entropy in excess of that acquired by
gravitational shock heating alone. Its temperature will therefore
exceed the virial temperature after infall and adiabatic
compression. As the level of preheating increases, gas at
progressively larger radii in the halo undergoes Jeans smoothing
effects. We now calculate this in detail.

We first construct the default entropy profile of the gas without
preheating. We assume that in the absence of heating or cooling
processes, the gas distribution traces that of the dark matter, an
ansatz which is indeed observed in numerical simulations (e.g.,
\cite{frenketal}) (the dark matter is assumed to follow the NFW
\cite{NFW} profile). This assumption becomes inaccurate at the very
center of the halo, where finite gas pressure causes the gas
distribution to be more flattened and less cuspy than the dark matter
density distribution.  In particular, even in the absence of
preheating the IGM has a finite temperature after decoupling from the
CMB and cooling adiabatically: $T_{\rm min}(z) \approx 2.73 (1+z_{d})
[(1+z)/(1+z_{d})]^{2}$, where the matter-radiation decoupling redshift
$z_{d} \approx 150$. This gives rise to a finite entropy floor:
\begin{equation}
K_{\rm min}= 4.6 \times 10^{-2} {\rm eV cm^{2}},
\label{eqn:min_entropy}
\end{equation}
independent of redshift. Thus, there will be a finite core in the gas
density profile even in the absence of preheating. In regions where
gas traces the dark matter, hydrostatic equilibrium gives the entropy
profile due to shock heating as:
\begin{eqnarray}
\label{eqn:entropy_shock}
&&K_{shock}(r)=\frac{1}{\rho_{g}(r)^{\gamma}} \times \\ \nonumber
&& \left[ \int_{r_{\rm vir}}^{r} - \frac{G
M(r) \rho_{g}(r)}{r^{2}}dr 
+  \frac{\rho_{g}(r_{\rm vir})}{\mu m_{B}} k_{B} T_{\rm vir}.
(r_{\rm vir}) \right].
\end{eqnarray}
The final entropy profile is therefore $K(r)={\rm max}[K_{\rm
min},K_{\rm shock}(r)]$. Note that the mean molecular weight is
$\mu=0.59$ for fully ionized gas and $\mu=1.22$ for fully neutral
primordial gas; in the paper we are dealing with the case where the
ionization fraction is small and therefore use the latter figure. This
results in virial temperatures which are higher by a factor $\sim 2$.

What is the effect of preheating on the entropy profile? If the infall
velocity does not exceed the local sound speed, then the gas is
accreted adiabatically and no shock occurs; the gas entropy $K_{\rm
IGM}$ is therefore conserved. If gas infall is supersonic, then the
gas is shocked to the entropy $K_{\rm shock}$ computed in equation
(\ref{eqn:entropy_shock}). \scite{tozzinorman} found that the
transition between the adiabatic accretion and shock heating regime is
very sharp. To a very good approximation $K(M)={\rm max}(K_{\rm
shock}(M),K_{\rm IGM})$; with this new entropy profile we can compute
density and temperature profiles. In fact, for most high redshift
mini--halos the 'preheating' entropy exceeds the shock entropy even at
the virial radius, so the gas mass is accreted isentropically. This
occurs when the entropy floor exceeds:
\begin{equation}
K_{\rm IGM} > 2.3 \, \, {\rm eV \, cm^{-3}} \left( \frac{T_{\rm vir}}{5000
  \, {\rm K}} \right) \left( \frac{\delta}{50} \right)^{-2/3}
  \redshifthi^{-2},  
\end{equation} 
where the gas overdensity $\delta \sim 50$ at the virial radius in the
absence of preheating is a weak function of the NFW concentration
parameter $c$. Although $c$ depends weakly on the collapse redshift,
for simplicity we shall assume in this paper that $c=5$ for all
mini--halos.

We have compared our entropy profiles calculated with
equation~(\ref{eqn:entropy_shock}) with the prescription in
\scite{tozzinorman} (and also adopted by \scite{oh_benson}). In this
method an accretion history for a halo is prescribed via extended
Press-Schechter theory; this allows one to compute the strength of the
accretion shock and thus the gas entropy (using standard
Rankine-Hugoniot jump conditions) for each Lagrangian mass shell. The
two methods agree extremely well; we therefore use
equation~(\ref{eqn:entropy_shock}) for both speed and simplicity.

Given an entropy profile, from hydrostatic equilibrium the density
profile of the gas is then given by:
\begin{eqnarray}
&&\rho(r)=\tilde{K}(r)^{-1/\gamma} \times \\ \nonumber &&\left[
P(r_{\rm vir})^{1-\frac{1}{\gamma}} + \frac{\gamma-1}{\gamma}
\int_{r_{\rm vir}}^{r} - dr^{\prime} \frac{G M(r^{\prime})}{r^{\prime
2}} \tilde{K}(r^{\prime})^{-1/\gamma} \right]^{1/(\gamma-1)}.
\end{eqnarray}
where $\tilde{K} \equiv P/\rho^{5/3} = k_{B} (\mu m_{H})^{-5/3}
K$. The temperature profile can then be determined from $T=\frac{\mu
m_{H}}{k_{B}} \tilde{K} \rho_{\rm gas}^{\gamma-1}$.

The solution of this equation requires a boundary condition which sets
the overall normalization, here expressed as $P(r_{\rm vir})$. There
is some ambiguity in this choice. The often used boundary condition
$M_{g}=f_{B} M_{halo}$ at $r=r_{\rm vir}$ is unphysical as it does not
take into account the suppression of accretion due to finite gas
entropy. We make the following choice. Let us define $P_{\rm shock}
(r_{\rm vir}) \equiv (\Omega_{b}/\Omega_{m}) \rho(r_{\rm vir})/(\mu
m_{H}) k_{B} T_{\rm vir}$, the pressure at the virial radius due to
shock heating alone. For $K_{\rm shock}(r_{\rm vir}) > K_{\rm IGM}$,
the final conditions at the virial radius are not strongly affected by
the entropy floor, since the shock boosts the gas onto a new
adiabat. We therefore set $P(r_{\rm vir})=P_{\rm shock} (r_{\rm
vir})$. The boundary condition must change when $K_{\rm IGM} > K_{\rm
shock}(r_{\rm vir})$, when accretion takes place isentropically, and
the entropy floor is fundamental in determining the gas pressure. In
this case, $P(r_{\rm vir})$ (which is essentially a constant of
integration) is chosen so that $\rho(r) \rightarrow \bar{\rho}$ as $r
\rightarrow \infty$. The latter boundary condition implicitly assumes
that hydrostatic equilibrium prevails beyond the virial radius. This
is questionable, but is arguably reasonable: the gas sound speed
$c_{s}=(\gamma K_{\rm IGM} \rho^{\gamma-1})^{1/2}$ in preheated gas is
much higher than in cold gas, which is essentially in free-fall. The
sound-crossing lengthscale over which hydrostatic equilibrium can be
established, $L_{sc} \approx c_{s} t_{H}$, is:
\begin{equation}
L_{sc} \approx 1 \, {\rm kpc} \, h^{-1} \left( \frac{{K}_{\rm IGM}}{3
    {\rm eV \, cm^{2}}} \right)^{1/2} \redshifthi^{-0.5} \left(
    \frac{\delta}{10} \right)^{1/3}
\end{equation} 
which is $\sim 5$ times larger than the virial radius for a $\sim
6000$K halo at the same redshift. Indeed, by definition $L_{sc} >
r_{\rm vir}$ for Jeans smoothing effects to be important.

\begin{figure}
\psfig{file=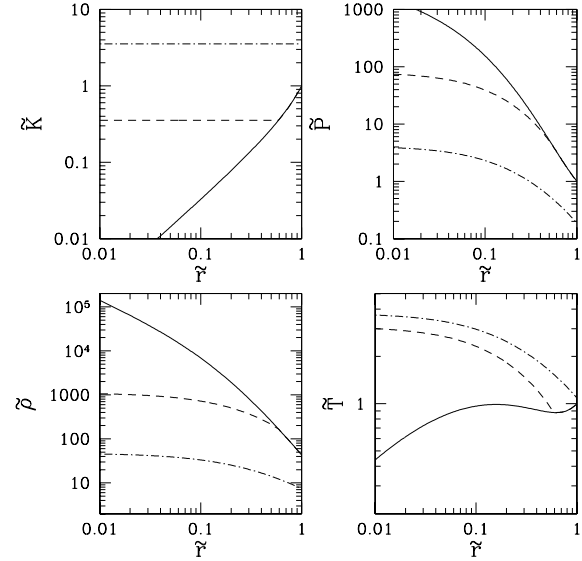,width=80mm}
\caption{The dimensionless entropy $\tilde{K}=K/K_{o}$, pressure
$\tilde{P}=P/P_{o}$, temperature $\tilde{T}=T/T_{o}$, and density
$\tilde{\rho}=\rho/\rho_{o}$, as a function of radius
$\tilde{r}=r/r_{\rm vir}$. $K_{o},P_{o},T_{o}$ are the values of these
quantities at $r_{\rm vir}$ without preheating, while
$\rho_{o}=\bar{\rho}_{b}$, the mean baryonic density. The entropy
profile $\tilde{K}(\tilde{r})$ uniquely specifies
$\tilde{P}(\tilde{r}),\tilde{T}(\tilde{r}),\tilde{\rho}(\tilde{r})$,
independent of halo mass or redshift.}
\label{fig:halo_profiles}
\end{figure}

Having established the appropriate boundary conditions, we can now
compute detailed gas profiles. The entropy, density, pressure and
temperature profiles as a function of $\tilde{r} \equiv r/r_{\rm vir}$
are shown in Fig \ref{fig:halo_profiles}. These profiles are of course
universal and independent of halo mass once $\hat{K}$ is set, if the
weak dependence of the NFW concentration parameter $c$ with mass is
ignored. As the entropy floor increases, the central pressure and
density decline, while the central temperature increases. We see that
reasonable values of the entropy floor $\hat{K} > 1$ produce dramatic
effects on the gas density profile of the halo, smoothing it out
considerably.

\begin{figure}
\psfig{file=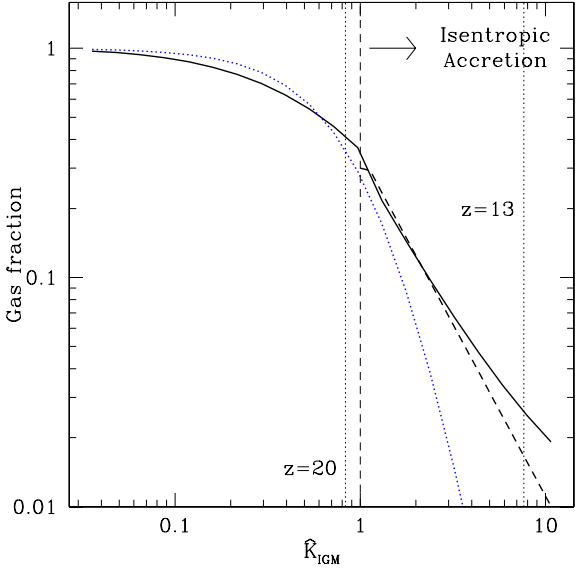,width=80mm}
\caption{The gas fraction within the virial radius $f_{\rm
gas}=(M_{g}/M_{\rm halo})/(\Omega_{b}/\Omega_{m})$ as a function of
the entropy parameter $\hat{K}$. The solid line indicates our fiducial
boundary conditions; the dotted line corresponds to the Gnedin (2000)
fit to numerical simulations; while the dashed line corresponds to the
Bondi accretion prediction (valid only for isentropic accretion,
$\hat{K} > 1$). All three show good agreement. The self-similarity of
this problem implies that the computed gas fractions applies to
mini--halos of all virial temperatures, provided $\hat{K}$ is
appropriately scaled. For illustrative purposes, appropriate values of
$\hat{K}$ for a mini--halo of $T_{\rm vir}=5000$K at $z=20,13$ are
shown (see Figure \ref{fig:Khat_z}).}
\label{fig:gas_frac}
\end{figure}

We now use these density profiles to compute how the accreted gas
fraction $f_{g} \equiv (M_{g}/M_{\rm halo})/(\Omega_{b}/\Omega_{m})$
scales with the entropy parameter $\hat{K}$. This is shown as the dark
solid line in Fig \ref{fig:gas_frac}.  Again, because of the
self-similarity of the problem, this plot is valid for mini--halos of
all virial temperatures at all redshifts, provided $\hat{K}$ is
appropriately re-scaled. It is reassuring to see that $f_{g}$ is
continuous at $\hat{K}=1$, when we switch from one boundary condition
to another. This need not have been the case, and gives us confidence
that we handle the transition to isentropic accretion correctly. We
see that realistic levels of the entropy floor (as computed in
\S\ref{section:entropy_floor}) causes a substantial depression in gas
fractions in mini--halos.

It is interesting to compare our derived gas fractions with other
estimates. For the case where $K_{\rm IGM} > K_{\rm shock}(r_{\rm
vir})$, accretion takes place isentropically. The halo will therefore
accrete gas at roughly the adiabatic Bondi accretion rate (e.g.,
\scite{baloghetal}):
\begin{equation}
\dot{M}_{b} \approx 1.86 \pi \lambda G^{2} M^{2}_{halo}
\tilde{K}_{IGM}^{-3/2}. 
\end{equation} 
The total accreted gas mass is then roughly $M_{gas} \approx {\rm min}
(f \dot{M}_{b} t_{H},(\Omega_{b}/\Omega_{m}) M_{halo})$, where $f$ is
some unknown normalization factor which takes into account the fact
that the total halo mass is not constant but was lower in the past
(and hence, that the gas accretion rate was lower in the
past). Another estimate which is a good fit to the results of
hydrodynamic simulations is \cite{gnedin00}:
\begin{equation}
\bar{M}_{b} \approx \frac{(\Omega_{b}/\Omega_{m})
M_{halo}}{[1+(2^{1/3}-1)M_{1/2}/M_{halo}]^{3}}.
\label{eqn:fgas_gnedin}
\end{equation}
There is only one free parameter: $M_{1/2}$, the mass of the halo in
which the gas mass fraction $f_{g}=0.5$. Gnedin (1999) shows that
$M_{1/2}$ is well approximated by the ``filtering mass''
$M_{F}$. However, $M_{F}$ depends on the unknown thermal history of
the IGM.  To make a self-consistent comparison, we compute $M_{1/2}$
with the density profiles computed with our fiducial boundary
conditions.  Interestingly, we find that $M_{1/2}$ roughly corresponds
to the halo mass when accretion begins to take place isentropically
$\hat{K} \sim 1$.

The results are shown in Figure \ref{fig:gas_frac}. All three
estimates agree well (note that the normalization $f$ of the Bondi
accretion prediction is a free parameter; the plot shown is for
$f=0.3$). The widely used \scite{gnedin00} fitting formula predicts
even lower gas fractions (and as we will see, clumping factors) at
high entropy levels $\hat{K} \gg 1$. On the other hand, the slope of
the $f_{g}(\hat{K})$ relation for our boundary conditions agrees very
well with the Bondi accretion prediction in this regime. Our boundary
conditions therefore yield fairly conservative estimates of the
effects of preheating. Moreover, unlike these other estimates, we are
able to compute detailed density profiles, which is crucial for some
of our later calculations.

\begin{figure}
\psfig{file=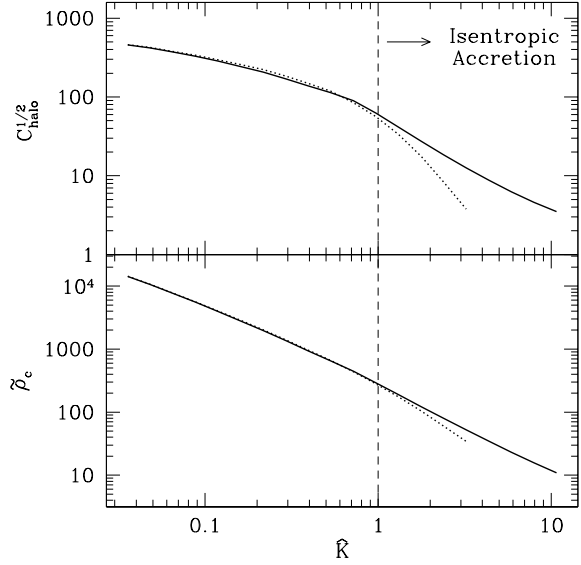,width=80mm}
\caption{The effect of increasing the entropy parameter $\hat{K}$ on
the central gas density $\tilde{\rho}_{c}=\rho_{c}/\bar{\rho}$ ({\it
bottom panel}), and gas clumping $C_{\rm halo}^{1/2}$ ({\it top
panel}), as defined in equation~(\ref{eqn:clumping_halo}). The solid
lines correspond to our fiducial boundary conditions, while the dashed
lines correspond to adjusting $P(r_{\rm vir})$ to reproduce the gas
fractions from the Gnedin (2000) fit to numerical simulations (the
results for the Bondi accretion prediction are almost identical to the
solid curve).  Both $\tilde{\rho}_{c}$ and $C_{\rm halo}^{1/2}$
decline rapidly as $\hat{K}$ increases. }
\label{fig:dual_clump_central}
\end{figure}

There are two other quantities which are of particular interest when
computing density profiles. One is the central density $\rho_{c}$,
which affects the ability of mini--halos to form $\htwo$ in the face
of UV photo-dissociation. Another is the gas clumping of the halo,
defined as:
\begin{equation}
C_{\rm halo}=\frac{\langle n^{2} \rangle}{\langle n \rangle^{2}}. 
\label{eqn:clumping_halo}
\end{equation}
where the brackets indicate a volume averaged quantity,
\begin{equation}
\langle X \rangle = \frac{1}{V} \int_{0}^{r_{\rm vir}} dr 4 \pi r^{2} X.
\end{equation}
 Note that $C_{\rm halo} \ge 1$ always. The clumping factor $C_{\rm
halo}$ plays a central role in determining the photon budget required
for reionization; we evaluate the global clumping factor in the
following section. In Figure \ref{fig:dual_clump_central}, we show the
effect of increasing the entropy parameter $\hat{K}$ on the central
density and clumping factor. Both decline rapidly with $\hat{K}$. We
shall use these two results in the following sections.

We now use these gas density profiles to compute
global effects of mini--halo suppression.

\section{Global Effects of Mini--Halo Suppression}
\label{section:global}

\subsection{Suppression of Collapsed Gas Fraction and Gas Clumping}
\label{subsection:coll_clump}

An entropy floor suppresses the fraction of gas which is bound within
mini--halos. As we discuss in \S\ref{subsection:21cm}, the mean 21cm
emission from mini--halos is directly proportional to this global gas
fraction; if it is strongly suppressed the signal will be
unobservable. The global collapsed gas fraction is:
\begin{equation}
f_{\rm gas} = \frac{1}{\rho_{b}} \int_{M_{l}}^{M_{u}} dM \frac{dn}{dM}
f_{b} f_{\rm halo,g}(\hat{K}) M
\end{equation}
where $f_{B}=\Omega_{b}/\Omega_{m}$, $f_{\rm halo,g}(\hat{K})$ is the
halo gas fraction as in Figure \ref{fig:gas_frac}, and $\rho_{b}$ is
the comoving baryon density. This is shown in the bottom panel of
Figure \ref{fig:dual_redshift_fc}. Curves are shown for no preheating
$K_{\rm IGM}=K_{\rm min}$, fixed values of the entropy floor $K_{\rm
IGM}=1,10 \, {\rm eV \, cm^{2}}$ (which correspond to the IGM settling
to some temperature at a given redshift and remaining at constant
entropy thereafter), and the redshift dependent entropy $K_{\rm
IGM}(z)$ shown in the top panel of Fig. 2, which reflects the
increasing entropy of gas which Compton cools at late epochs. For
realistic values of $K_{\rm IGM}$, the collapsed gas fraction is
suppressed by 1--2 orders of magnitude.

\begin{figure}
\psfig{file=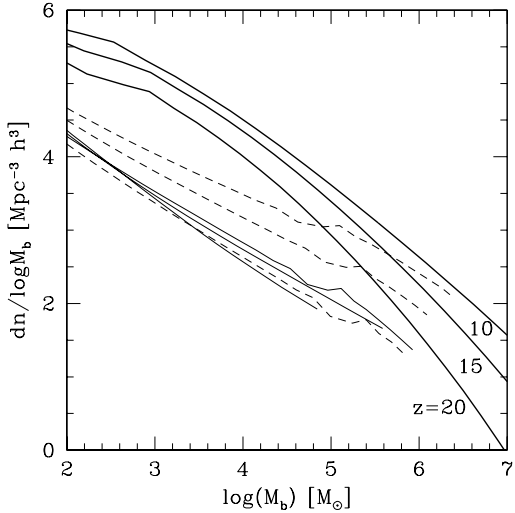,width=80mm}
\caption{The mass function of baryons, as in equation
\ref{eqn:baryon_mf}, for $z=10,15,20$. Dark solid curves show the mass
function for $K_{\rm IGM}=K_{\rm min}$, light solid curves are for
redshift-dependent entropy $K_{\rm IGM}(z)$ as in the top panel of
Figure 2, and dashed curves are for a fixed entropy floor $K_{\rm
IGM}=3 \, {\rm eV \, cm^{2}}$. Note that the baryonic mass function is
very similar at all redshifts for $K_{\rm IGM}(z)$. For $T_{\rm vir} >
10^{4}$K, an entropy floor has no effect on the mass function, which
reverts back to the dark solid curves (we do not extend the light
solid and dashed curves above the corresponding halo masses).}
\label{fig:plot_mass_function}
\end{figure}

\begin{figure}
\psfig{file=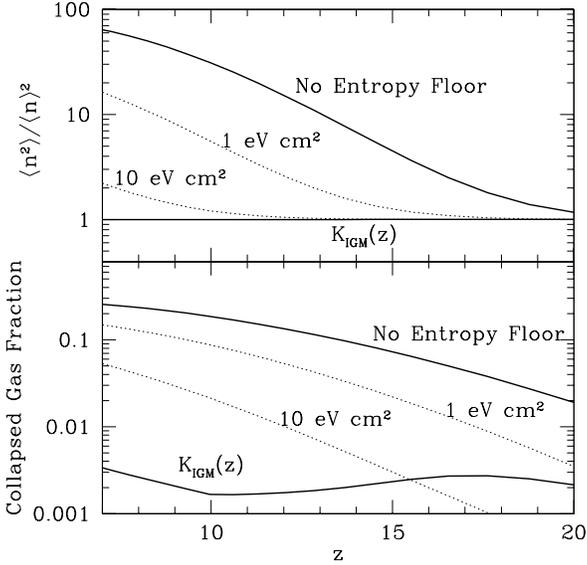,width=80mm}
\caption{The evolution of the collapsed gas fraction in mini--halos
({\it lower panel}) and the gas clumping factor ({\it upper panel})
with redshift. The results are shown for no preheating (with $K_{\rm
IGM}=K_{\rm min}$), as well as constant values of entropy $K_{\rm
IGM}=1,10 \, {\rm eV \, cm^{2}}$ and the redshift-dependent entropy
$K_{\rm IGM}(z)$ shown in the top panel of Figure 2.  Both the global
gas fraction in mini--halos and gas clumping are strongly suppressed
for realistic values of the entropy floor. In particular, $C
\rightarrow 1$ for realistic values of $K_{\rm IGM}$, greatly reducing
the photon budget required for reionization.}
\label{fig:dual_redshift_fc}
\end{figure}

It is interesting to plot the effect of preheating on the baryonic
mass function. This can be computed simply as:
\begin{equation}
\frac{dN}{dM_{b}}=\frac{dN}{dM_{h}}\times \frac{dM_{h}}{dM_{b}}.
\label{eqn:baryon_mf}
\end{equation}
We show this in Fig. \ref{fig:plot_mass_function}, at $z=10,15,20$ for
$K_{\rm IGM}(z)$ as in the top panel of Fig. 2, and for a fixed
entropy floor $K_{\rm IGM}=3 \, {\rm eV \, cm^{2}}$. As expected, the
effect of preheating is most drastic at low masses. Note that the
baryonic mass function is very similar at all redshifts for $K_{\rm
IGM}(z)$; this can be also seen in the bottom panel of
\ref{fig:dual_redshift_fc}, where the collapsed fraction in
mini--halos does not change appreciably with redshift.

Another quantity of great interest is the global gas clumping factor
$C=\langle n^{2} \rangle/\langle n \rangle^{2}$. Gas clumping shortens the
recombination time $t_{rec} \approx 1/(\alpha n C)$ and thus increases
the total number of photons per baryon required to achieve
reionization. It is given by:
\begin{equation}
C= (1-f_{V})C_{\rm IGM} +  \int_{M_{l}}^{M_{u}} \left(\frac{t_{\rm
  evap}}{t_{H}}\right) C_{\rm
  halo}(\hat{K}) \frac{dn}{dM} V_{\rm halo} dM
\label{eqn:clumping}  
\end{equation}
where $f_{V}$ is the collapsed fraction by volume of mini--halos,
$C_{\rm IGM} \approx 1$ is the clumping factor of the IGM, $C_{\rm
halo}(\hat{K})$ is the halo clumping factor as calculated in the
previous section and shown in the top panel of Figure
\ref{fig:dual_clump_central}, and $V_{\rm halo}=(4\pi/3) r_{\rm
vir}^{3}$ is the volume of a mini--halo. The factor $(t_{\rm
evap}/t_{H})$ deserves some explanation. Mini--halos only contribute
to recombinations when they are photoionized. However, once they are
exposed to ionizing radiation, they will be evaporated, essentially in
the sound crossing time of photo-ionized gas
\cite{shapiro98,barkanaloeb99}.  As in \scite{haimanabelmadau}, we
therefore set the evaporation time $t_{\rm evap}=r_{\rm vir}/(10 {\rm
\, km \, s^{-1}})$, and weight the halo clumping factor by the duty
cycle $t_{\rm evap}/t_{H}$.

The clumping factor in equation \ref{eqn:clumping} is a lower limit to
the total gas clumping; there will be an additional contribution from
larger scales as well. However, the clumping due to minihalos is
dominant at the high redshifts of interests.

Our result is shown in the top panel of Figure
\ref{fig:dual_redshift_fc}. For the no entropy floor $K_{IGM}=K_{\rm
min}$ case (top curve), gas clumping is much lower at high redshift,
since fewer mini--halos have collapsed. Once a patch of IGM is
reionized early at high redshift, then an entropy floor is established
and subsequent gas clumping is suppressed. Thus, early star formation
{\it reduces} the total photon budget required to achieve full
reionization. We see that gas clumping is reduced by an order of
magnitude even for the low entropy levels $\sim 1 {\rm eV \, cm^{2}}$
associated with reionization at high redshift $z\sim 20$. For the
redshift-dependent entropy $K_{\rm IGM}(z)$ shown in the top panel of
Figure \ref{fig:Khat_z}, mini--halo clumping is strongly suppressed,
and $C \approx 1$.

\subsection{Suppression of $\htwo$ formation}
\label{subsection:h2_metals}

${\rm H_{2}}$ formation and cooling in mini--halos will be suppressed
relative to the no preheating case. This is because the finite entropy
of the gas allows it to resist compression; thus, the gas is at
considerably lower density. Collisional processes with cooling times
$t_{cool} \propto 1/n$ will be suppressed relative to radiative
processes which photo-dissociate $\htwo$. It has been recognized that
gas profiles should have a central core rather than a cusp (e.g.,
\scite{shapiro99}) and that lower central densities will affect
$\htwo$ chemistry \cite{hrl96,tegmark}. However, the fact that
preheating can produce a much larger, lower density core than hitherto
considered, and the subsequent implications for $\htwo$ formation and
cooling, has not been explored.

It has been shown that $\htwo$ formation is enhanced in fossil HII
regions due to increased electron fraction there
\cite{ricotti}. Although $\htwo$ formation is very slow in the low
density IGM, a high initial electron abundance will persist when
weakly ionized gas is accreted onto non-linear structures, and aid
$\htwo$ formation at that point \cite{oh2000}.  We can conservatively
take into account the $\htwo$ enhancement by assuming the maximum
initial abundance of $\htwo$, $x_{\htwo} \sim 10^{-3}$; this hard
upper limit is independent of density or ionization fraction and is
due to 'freeze--out' (\scite{oh_haiman}, hereafter OH02).  The $\htwo$
abundance can only exceed this if three-body processes are important,
which only takes place at very high density $n > 10^{8} {\rm
cm^{-3}}$.

In OH02, we showed that the minimum temperature $T_{\rm min}$ a parcel
of gas can cool to depends almost exclusively on $t_{\rm cool}/t_{\rm
diss} \propto J_{\rm UV}/n$ (see Figure 6 of OH02). This scaling
behavior is only broken when the gas approaches high densities $n >
10^{4} {\rm cm^{-3}}$ (at this point, the cooling time becomes
independent of density and $t_{\rm cool}/t_{\rm diss} \propto J_{\rm
UV}$). However, the latter regime is never reached in pre-heated gas,
which is at much lower densities ($n \sim 10^{4} {\rm cm^{-3}}$
corresponds to $\delta \sim 6 \times 10^{6}$ at $z=19$). In subsequent
discussion, we shall assume the dependence of $T_{\rm min}$ on $J_{\rm
UV}/n$ shown in Figure 6 of OH02.

\begin{figure}
\psfig{file=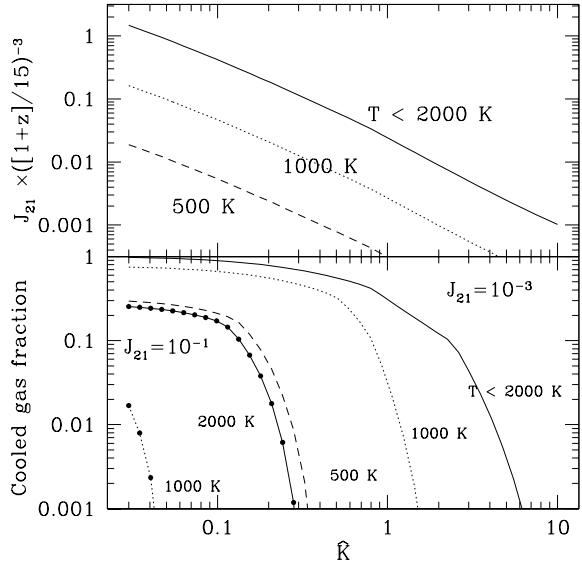,width=80mm}
\caption{{\it Top panel:} The critical radiation field $J_{\rm UV}$
required to suppress gas cooling below some temperature $T$, as a
function of entropy parameter $\hat{K}$. An entropy floor greatly
reduces the radiation field required to photo-dissociate $\htwo$, by
several orders of magnitude. Note that $J_{\rm UV} \propto (1+z)^{3}$;
the curves shown are for a halo at $z=14$. {\it Bottom panel:} The
fraction of gas in a halo able to cool below temperature $T$, for
$J_{\rm UV}=10^{-3}$ (lines), and $J_{\rm UV}=10^{-1}$ (dots connected
by lines), at $z=14$. Beyond a critical value of $\hat{K}$, the cooled
gas fraction plummets dramatically.}
\label{fig:dual_H2_Khat}
\end{figure}

Thus, in the presence of a radiation field $J_{\rm UV}$, the gas must
be at a minimum density $n_{\rm crit}$ to cool down to a temperature
$T_{\rm min}$. We have already calculated the maximum central density
$n_{c}$ of gas in a halo given an entropy parameter $\hat{K}$, as in
Figure \ref{fig:dual_clump_central}. If the central density is less
than the critical density, $n_{c} < n_{\rm crit}$, then {\it none} of
the gas in the halo can cool down to $T_{\rm min}$. For a given
entropy parameter $\hat{K}$, there is therefore a minimum radiation
field $J_{\rm UV}$ above which no gas can cool down to $T_{\rm
min}$. We plot this in the top panel of Figure
\ref{fig:dual_H2_Khat}. This plot is valid for all halos at all
redshifts provided $\hat{K}$ and $J_{\rm UV}$ are both rescaled
appropriately (note that since $n \propto (1+z)^{3}$, to keep $J_{\rm
UV}/n$ constant, $J_{\rm UV} \propto (1+z)^{3}$). To get a sense of
typical values of $J_{\rm UV}$, the radiation field corresponding to
$n_{\gamma}$ ionizing photons per baryon in the universe is $J_{\rm
UV} \approx \frac{h_{P} c}{ 4 \pi} n_{\gamma} n_{b} (1+z)^{3} /
10^{-21} {\rm erg \, s^{-1} \, cm^{-2} \, Hz^{-1} \, sr^{-1}} \approx
10 n_{\gamma} \left(\frac{1+z}{16}\right)^{3}$ (where $n_{b}$ is the
comoving baryon number density, and $h_{P}$ is the Planck constant).

We see that an entropy floor greatly reduces the radiation field
required to prevent gas cooling. For reasonable values of $\hat{K}$,
the reduction can be as much as four orders of magnitude. Thus, even
if there is only a very weak radiation field, in the presence of an
entropy floor effective $\htwo$ cooling and star formation will be
quenched. Note also that the cooling time exceeds the Hubble time for
large values of $\hat{K}$. This regime is reached only when gas
densities are so low that even a very weak radiation field can
dissociate $\htwo$: $t_{\rm cool} \sim t_{\rm diss} \sim 2 \times
10^{8} (J_{\rm UV}/10^{-4})^{-1}$yr.

Without an entropy floor, the density profiles of mini--halos at a
given redshift are self-similar; thus, roughly comparable fractions of
their gas can cool down to $T_{\rm min}$ and will be available as fuel
for star formation. In the presence of an entropy floor, the fraction
of gas with $n > n_{\rm crit}$ is greatly reduced, and much less gas
can cool. Moreover, the self-similarity is broken: shallower potential
wells (which have higher $\hat{K}$) are more strongly affected by an
entropy floor, and their central gas densities are much lower. Using
our halo density profiles, we can calculate the fraction of gas above
the critical density, $f_{gas}= M_{b}(> n_{\rm
crit})/[(\Omega_{b}/\Omega_{m}) M_{\rm halo}]$. We plot this in the
lower panel of Fig \ref{fig:dual_H2_Khat}. At some critical value of
$\hat{K}$, the fraction of gas which can cool plummets
dramatically. This value of $\hat{K}$ corresponds roughly to when the
low density core falls below the critical density.

Note that we assume static rather than evolving density profiles.  In
reality, as gas cools, densities should increase as the gas loses
pressure support, making the gas more resistant to $\htwo$
photo-dissociation (Machacek et al. 2001). However, the dynamical time
$t_{\rm dyn} \approx 10^{8} ({\rm n/1 \, cm^{-3}})^{-1/2}$yr is
typically much longer than the photo-dissociation time $t_{diss}
\approx 2 \times 10^{4} J_{\rm UV}^{-1} f_{\rm shield}^{-1}$yr (where
the factor $f_{\rm shield}$ takes into account self-shielding; see
discussion in section 4.2.3 of OH02, and references therein). So our
use of the initial density profile to estimate timescales should be
approximately valid.

\begin{figure}
\psfig{file=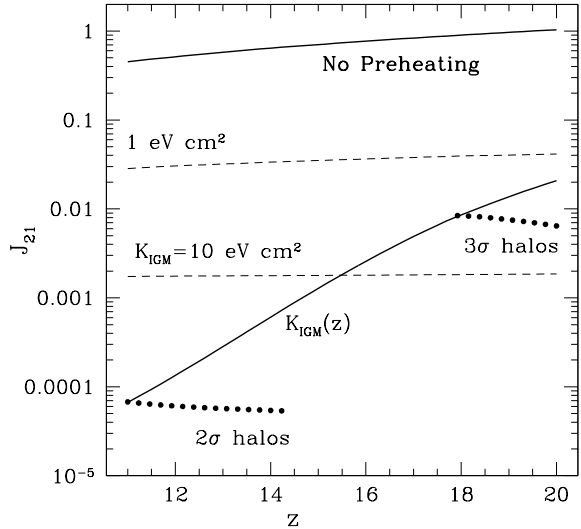,width=80mm}
\caption{The critical radiation field $J_{\rm UV}$ required to
suppress gas cooling below $T=1000$K, as a function of redshift z. The
curves are shown for a $T_{\rm vir}=8000$K halo, which is an upper
limit in mass to the mini--halo population: smaller mini--halos will
require a weaker radiation field for cooling to be suppressed. The
curves are shown for four different entropy floors: no preheating
$K_{\rm IGM}=K_{\rm min}$, an evolving entropy $K_{\rm IGM}(z)$ as
shown in Fig. 2, and fixed entropy floors $K_{\rm IGM}=1,10 \, {\rm eV
\, cm^{2}}$. Points are also shown for $2\sigma$ and $3\sigma$ halos
at each redshift, assuming $K_{\rm IGM}(z)$ (the points end when
$T_{\rm vir} < 1000$K and $T_{\rm vir} > 8000$K respectively). Typical
entropy floors at each redshift reduce the critical radiation field
required for $\htwo$ suppression by $2-4$ orders of magnitude,
compared to the no preheating case.}
\label{fig:Jcrit_redshift}
\end{figure}

We can also compute the critical radiation field required to suppress
$\htwo$ cooling down to some temperature $T$ as a function of
redshift. We plot this in Fig. \ref{fig:Jcrit_redshift} for a $T_{\rm
vir}=8000$K halo; such a halo represents an upper mass limit to the
mini--halo population. Smaller halos, which have lower central
densities for a given entropy floor, will require a weaker radiation
field for $\htwo$ cooling to be suppressed. This is shown for four
different entropy floors: $K_{\rm IGM}(z)$ as shown in Fig. 2, no
preheating $K_{\rm IGM}=K_{\rm min}$, and fixed entropy floors $K_{\rm
IGM}=1,10 \, {\rm eV \, cm^{2}}$. Gas cooling in fossil $\htwo$
regions is very easily suppressed; the radiation field required is 2-4
orders of magnitude lower than required in the no preheating case.

One possible caveat is that since fossils are likely to be
metal-enriched, gas cooling is dominated by metal line cooling, rather
than $\htwo$ cooling. We argue that this is unlikely, for two reasons:
1) Metal line cooling only dominates $\htwo$ cooling above a critical
metallicity $Z > 10^{-3} Z_{\odot}$ \cite{bromm01,hellsten}. This
level of metal enrichment corresponds to $\sim 1 f_{Z}^{-1}$ ionizing
photon per baryon in the universe (where $f_{Z}$ is the volume filling
factor of metals). Thus, metal line cooling becomes significant only
at late times. 2) After an ionizing source turns off and explodes as a
supernova, the metal-polluted region is much smaller than the fossil
HII region \cite{madau01}. Most of the volume is therefore still of
pristine composition, and undergoes the entropy floor suppression we
have described. The metal-polluted region lies close to the high
density peak where the very first stars formed, where in any case,
more massive halos $T_{\rm vir} > 10^{4}$K will collapse. To
summarize: metal line cooling is therefore unlikely to spoil our
assumption of adiabaticity in most of the volume of the fossil HII
region. The exception is at the most highly biased density peaks,
where metals can be thought of as effectively increasing the star
formation efficiency.

The net result is that entropy injection greatly boosts the negative
feedback from early star formation: the entropy floor in reionized
regions results in low density cores in the center of halos, in which
$\htwo$ is easily photo-dissociated by a weak external UV radiation
field.
 
\subsection{Negative Feedback from X-rays}

It is often argued that X-rays boost $\htwo$ production and cooling in
mini--halos, by penetrating deep into the dense core and increasing
the free electron fraction, which is critical for gas phase $\htwo$
production. Thus, X-rays are thought to exert a positive feedback
effect, counter-acting photo-dissociation by UV radiation
\cite{hrl96,haimanh2}. We show here by fairly general arguments that
X-rays in fact exert a strong {\it negative} feedback effect, due to
the entropy that they inject into the IGM. This prevents gas from
compressing to sufficiently high density to produce $\htwo$ and cool
efficiently, and far outweighs any positive feedback from increasing
the free electron fraction.

For the sake of definiteness, we consider an intrinsic source spectrum
of the form
\begin{eqnarray}
J=&& J_{\rm UV} \left(\frac{\nu}{\nu_{L}} \right)^{-1} \,\, h\nu
< 13.6 {\rm eV} \\
=&& J_{\rm X} \left(\frac{\nu}{\nu_{L}} \right)^{-1} \,\, 13.6 {\rm eV} <
h\nu < 10 {\rm keV}  \nonumber
\end{eqnarray}
where $h \nu_{L}= 13.6$eV and $J_{\rm UV}/J_{\rm X}=f_{\rm
break}=1-100$ is the spectral break at the hydrogen Lyman edge; note
that $f_{break} \ge 1$ always. A spectrum of the form $J_{\nu} \propto
\nu^{-1}$ is characteristic of the mean spectra of quasars
\cite{elvis94}, as well as inverse Compton emission from high-redshift
SN \cite{oh2001}. It is distinctive in that $\nu J_{\nu}$=const, i.e.,
there is equal power per logarithmic interval (though our arguments
can be generalized to other spectra). Note that the actual ionizing
spectrum at any given point in the IGM will be much harder, since
photoelectric absorption hardens the spectrum away from the source.

What is the heating associated with this radiation field? As we argued
in \S\ref{section:X_ray_floor} above, all energy in the X-ray
radiation field with ${\rm E_{L} < E < E_{\rm thick}}$ (as defined in
\S\ref{eqn:Ethick}) will be absorbed by the IGM, since the universe is
optically thick at these frequencies. The radiation field may be
subdivided into two components. The ``mean field'' consists of photons
with $E > E_{\rm overlap}$, where $E_{\rm overlap}$ is defined by:
\begin{equation}
\lambda_{\rm mfp} (E) > n_{\rm source}^{-1/3} \ \ {\rm for} \ \ E >
E_{\rm overlap}   
\end{equation}
i.e., photons with $E > E_{\rm overlap}$ (typically, $E_{\rm overlap}
\sim 100$eV) have a mean free path greater than the mean separation
between sources, so that a homogeneous ionizing background is
established. The ``fluctuating field'' consists of photons with $E <
E_{\rm overlap}$; this component is dominated by radiation from a
single source, and is subject to large Poisson fluctuations. We are
interested in the heating due to the 'mean field', which can lead to
an entropy floor even outside the HII regions of ionizing sources.  It
is:
\begin{equation}
\dot{E}_{\rm heat} = f_{\rm heat} \int_{\nu_{\rm overlap}}^{\nu_{\rm
thick}} d\nu \epsilon_{\nu} \approx f_{\rm heat} \frac{4 \pi}{l_{H}}
J_{\rm X} \nu_{L} {\rm ln} \left( \frac{\nu_{\rm thick}}{\nu_{\rm
thin}} \right)
\end{equation}
where $f_{\rm heat}$ is the fraction of energy of the hot
photo-electron created by an X-ray which goes into heating. We have
conservatively set $\epsilon_{\nu} \approx (4 \pi J_{\nu})/l_{H}$;
this underestimates $\epsilon_{\nu}$ for a given $J_{\nu}$. For the
spectrum we have chosen, the result is only logarithmically sensitive
to the integration bounds; we set ${\rm ln} \left( {\nu_{\rm
thick}}/{\nu_{\rm thin}} \right) \approx 2.3$. For steeper spectra,
$J_{\nu} \propto \nu^{-\alpha}$ where $\alpha > 1$, the lower
integration limit becomes important.

Over a Hubble time $t_{H}$, the internal energy density $U=n k_{B} T
\approx \dot{E}_{\rm heat} t_{H}$
of the gas becomes:
\begin{equation}
U \approx f_{\rm heat} \frac{4 \pi}{c} J_{\rm X} \nu_{L} {\rm ln}
\left( \frac{\nu_{\rm thick}}{\nu_{\rm thin}} \right) 
\end{equation}
This make sense: since some fraction $f_{\rm heat}$ of the radiation
field goes directly into heat, $f_{\rm heat} 4 \pi (\nu J_{\nu})
\approx U c$. Thus, X-rays will heat the IGM to a temperature:
\begin{equation}
T=170 {\rm K} \left( \frac{f_{\rm heat}}{0.5} \right) \left(
\frac{J_{\rm X}} {10^{-2}} \right) \left( \frac{1+z}{15} \right)^{-3}.
\end{equation}
There is therefore a direct relation between the X-ray radiation field
and the entropy floor:
\begin{equation}
K_{\rm IGM}=1.9 \, {\rm eV \, cm^{2}} \left( \frac{f_{\rm heat}}{0.5}
\right) \left( \frac{J_{\rm X}} {10^{-2}} \right) \left( \frac{1+z}{15}
\right)^{-5}.
\label{eqn:Xray_entropy}
\end{equation}

As we argued in \S\ref{section:X_ray_floor} above, the IGM evolves
adiabatically for $x_{e} < 0.1 [(1+z)/15]^{2.5}$. What level of the
radiation field does this correspond to? For $x_{e} \le 0.1$, about
$f_{\rm ion} \sim 1/3$ of the injected energy goes toward performing
ionizations; the number of photo-ionizations is simply $E_{\rm
photon}/(37 \, {\rm eV})$. Thus, we obtain:
\begin{equation}
x_{e}=2.3 \times 10^{-4}  \left( \frac{f_{\rm ion}}{0.3} \right) \left(
\frac{J_{\rm X}} {10^{-2}} \right)  
\end{equation}
and so for $J_{\rm X} < 5$, the IGM evolves adiabatically (note also
that since $t_{rec} \approx 10  t_{H} (x_{e}/10^{-3})^{-1} (T/100 \, {\rm
K})^{-0.7} ([1+z]/15)^{-1.5}$, we can ignore recombinations). 

\begin{figure}
\psfig{file=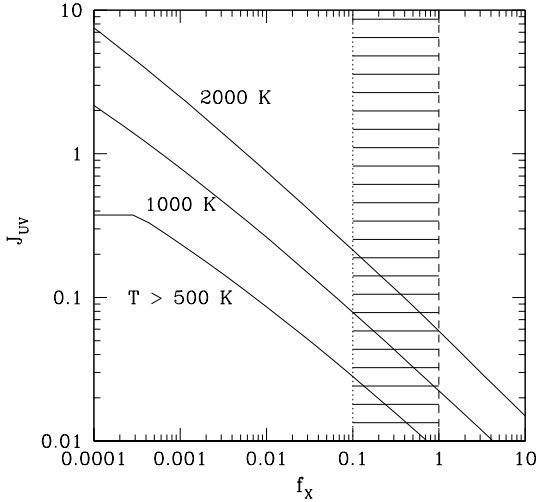,width=80mm}
\caption{The UV background $J_{\rm UV}$ required to suppress cooling
down to some temperature $T_{\rm min}$, as a function of $f_{\rm
X}=J_{\rm X}/J_{\rm UV}$. The curves plotted are for a $T_{\rm
vir}=5000$K halo at $z=19$; the results may be rescaled to other halos
by rescaling the x-axis: $f_{X} \propto T_{\rm vir} (1+z)^{5}$. Also
shown as dashed and dotted lines are the maximum value of $f_{X}
\approx 1$ in the hardest sources, and the minimum typical value of
$f_{\rm X} \approx 0.1$ required for X-rays to exert a positive
feedback effect; the shaded region therefore shows the possible region
whereby X-rays were assumed exert a positive feedback effect. Instead,
we see that such a strong X-ray background exerts a {\it negative}
feedback effect. By inducing a low density core in the halo, an X-ray
background in the required zone reduces by $1-2$ orders of magnitude
the minimum UV flux required to suppress cooling, compared to the case
where $f_{X} \rightarrow 0$.}
\label{fig:Xray_J21}
\end{figure}

Given the entropy floor from equation \ref{eqn:Xray_entropy}, we can
now calculate density profiles of halos. In particular, we can
calculate the maximum central density of a halo from the entropy floor
induced by a given X-ray background. As before, from the central
density of a halo we can calculate the critical level of the LW flux
$J_{\rm UV}$ required to suppress $\htwo$ cooling down to some
temperature $T_{\rm min}$. This critical flux is shown as a function
of $f_{\rm X}=J_{\rm X}/J_{\rm UV}$ in Fig \ref{fig:Xray_J21}, for a
halo of $T_{\rm vir}=5000$K at $z=19$. The curves may be simply
rescaled for halos of different mass at different redshifts by
translating the x-axis: $f_{X} \propto T_{\rm vir} (1+z)^{5}$. The
effect of X-rays increases for less massive halos (which are less able
to withstand an entropy floor) at lower redshifts (when gas densities
are lower). Also shown are the maximum value of the X-ray background
$f_{\rm X} \approx 1$ for the hardest sources, as well as the typical
minimum value of the X-ray background required for positive feedback
to be effective $f_{\rm X} \approx 0.1$ \cite{haimanh2}.  The shaded
region therefore denotes the X-ray fluxes thought to produce positive
feedback effects. We see for these large values of the X-ray
background, X-rays {\it cannot} exert a positive feedback effect. The
required X-ray background is so large that it would produce an
unacceptably large entropy floor, which prevents the formation of
dense regions where $\htwo$ can withstand photodissociation. Instead,
X-rays produce an over-all negative feedback effect. In fact, values
of $f_{X} \approx 0.1-1$ reduce by almost two orders of magnitude the
value of the UV background required to photo-dissociate $\htwo$ and
prevent it from cooling down to low temperatures.

The only way of circumventing these difficulties is if X-rays turn on
suddenly at some epoch. Halos which have already collapsed will be
unaffected by the entropy floor, but will undergo the usual positive
feedback from X-rays. However, the subsequent generation of mini--halos
will be suppressed by the entropy floor.

\subsection{21cm Observational Signatures}
\label{subsection:21cm}

At present, the only observational probes proposed for small-scale
structure at high redshift such as mini--halos are 21cm observations
with future radio telescopes such as the Square Kilometer Array
(SKA)\footnote{see http://www.nfra.nl/skai} and the Low Frequency
Array (LOFAR)\footnote{see http://www.astron.nl/lofar}. It has been
proposed that mini--halos could be detected statistically in emission
\cite{mh1,mh2} and individually in absorption along the line of sight
to a high-redshift radio source \cite{furlanetto}. If an entropy floor
exists, such signatures will be strongly suppressed. 21cm signatures
of mini--halos therefore provide an interesting indirect probe of
reionization history: if mini--halos {\it are} detected in large
numbers at a given redshift, it would imply that the filling factor of
fossil HII regions is still small at that epoch. In particular, a
comoving patch where mini--halos are seen has likely never been
ionized before (which is a much stronger constraint than merely being
neutral at the observed epoch). This could prove a very useful probe
of the topology of reionization.

Mini--halos are too faint to be seen individually in emission, and can
only be detected statistically through brightness temperature
fluctuations. Provided $T_{\rm S} \gg T_{\rm CMB}$ (where $T_{\rm S}$
is the spin temperature), the 21cm flux $S$ is independent of the spin
temperature, and depends only on the HI mass, $S \propto
M_{HI}$. Thus, $S({\rm halos})/S({\rm IGM})$ is simply equal to the
collapsed gas mass fraction in mini--halos, which is always less than
unity. The mini--halo signal is much smaller if an entropy floor
exists, since the collapsed gas fraction declines rapidly: the bottom
panel of Figure \ref{fig:dual_redshift_fc} can simply be read off as
$S({\rm halos})/S({\rm IGM})$. Thus, the IGM dominates 21cm
emission. Mini--halos dominate only when: (i) the IGM spin temperature
has not yet decoupled from the CMB (the critical thermalization
radiation field $J_{21}$ such that $T_{S} \rightarrow
T_{\alpha}\approx T_{K}$, where $T_{\alpha}$ is the color temperature
of the radiation field, is $J_{21} \approx 5 ([1+z]/10)$ \cite{MMR})
(ii) the filling factor of fossil HII regions is small. The high {\it
WMAP} optical depth, which implies significant reionization at high
redshift, pushes the latter constraint to high redshift and thus low
radio frequencies $v_{\rm obs}= 93 ([1+z]/15)$MHz, making this a very
challenging observation.

We now turn to absorption signatures. How does the mini--halo 21cm
absorption signature scale with the IGM entropy? The IGM 21cm optical
depth is:
\begin{equation}
\tau_{\nu} \approx 10^{-2} \left( \frac{T_{\rm CMB}(z)}{T_{S}} \right)
\redshiftnine^{1/2} x_{\rm HI}.
\end{equation}
See \scite{MMR} for expressions for the spin temperature
$T_{S}(J_{21},n_{HI},T_{K})$; we do not reproduce them here. We use a
fit to the results of \scite{allison} for the collisional coupling
coefficient $C_{10}$ for $T_{K} < 1000$K, with a $C_{10} \propto
T_{K}^{-0.33}$ extrapolation for higher temperatures. The mini--halo
21cm optical depth along an impact parameter $\alpha$ to the halo
center is \cite{furlanetto}:
\begin{equation}
\tau_{\rm halo}(\alpha,\nu)= \frac{3 h_{P} c^{3} A_{10}}{32 \pi k_{B}
\nu_{o}^{2}} \int dR \frac{n_{HI}(r)}{T_{s}(r) \sqrt{\pi} b(r)} \\
\nonumber {\rm exp} \left[ - \frac{v(\nu)^{2}}{b^{2}(r)} \right]
\end{equation}
where $A_{10}=2.85 \times 10^{-15} {\rm s^{-1}}$ is the spontaneous
emission coefficient, $b(r)^{2}=2 k_{B}T_{K}/m_{p}$ is the gas Doppler
parameter, $r^{2}=R^{2}+\alpha^{2}$, and
$v(\nu)=c(\nu-\nu_{o})/\nu_{o})$. Unlike \scite{furlanetto}, we do not
attempt to model the velocity field of the infall region, which we
ignore; the total observed optical depth is taken to be $\tau_{\rm
tot}= \tau_{\rm halo} + \tau_{\rm IGM}$. The velocity field in the
preheated gas will differ significantly from the Bertschinger
self-similar solution they assume, since gas pressure retards
accretion; the infall region optical depth will be smaller because of
the reduced gas column. Since even in their case the contribution of
the infall region to 21cm absorption is small (see their Fig. 2, where
the equivalent width drops drastically for $\alpha > r_{\rm shock}$),
our neglect is justified.

\begin{figure}
\psfig{file=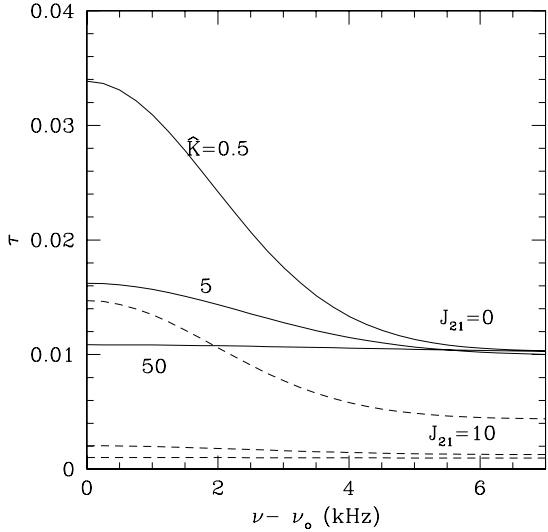,width=80mm}
\caption{Mini--halo optical depth profiles as a function of observed
frequency for a $2 \times 10^{6} {\rm M_{\odot}} h^{-1}$ halo at
$z=10$, all for impact parameters of $\alpha=0.3 r_{\rm vir}$. The
values of $\hat{K}=0.5,5,50$ shown correspond to IGM temperatures of
$T_{\rm IGM}=70,700,7000$K; since fossil HII regions at z=10 cannot
significantly cool below $7000$K, the value of $\hat{K}=50$ is most
appropriate. Solid lines correspond to $J_{21}=0$; dashed lines
correspond to $J_{21}=10$. The entropy floor makes the mini--halo
contribution to the 21cm absorption undetectable.}
\label{fig:plot_21cm_profile}
\end{figure}

In Fig. \ref{fig:plot_21cm_profile}, we show the optical depth as a
function of observed frequency, for different levels of the entropy
parameter $\hat{K}$. The values of $\hat{K}$ shown correspond to IGM
temperatures of $T_{\rm IGM}=70,700,7000$K; since fossil HII regions
at z=10 cannot significantly cool below $7000$K (see
Fig. \ref{fig:temp_params}), the value of $\hat{K}=50$ is most
appropriate. We see that the optical depth contribution from
mini--halos falls drastically with increasing $\hat{K}$. Also shown is
the optical depth for $J_{21}=10$. The radiation field drives $T_{S}
\rightarrow T_{K}$, reducing the 21cm optical depth $\tau \propto
T_{S}^{-1}$.

\begin{figure}
\psfig{file=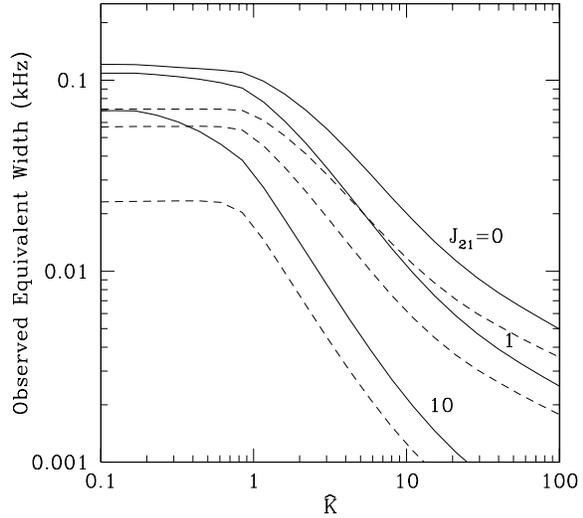,width=80mm}
\caption{The observed 21cm absorption equivalent width of a $2 \times
  10^{6} {\rm M_{\odot}} h^{-1}$ halo at $z=10$, for $J_{21}=0,1,10$,
  as a function of $\hat{K}$. Solid lines are for
  an impact parameter of $\alpha=0.3 r_{\rm vir}$, while dashed lines
  are for $\alpha=0.7 r_{\rm vir}$. The equivalent width falls rapidly
 as the entropy floor rises; for the expected value of
  $\hat{K}=50$ at this redshift, equivalent widths are very small.}
\label{fig:EW_21cm}
\end{figure}

Since $\tau_{\rm IGM}$ also depends on $\hat{K}$, the most relevant
quantity for detecting a mini--halo is the observed equivalent width:
\begin{equation}
\langle \Delta \nu \rangle = \frac{2}{1+z_{h}} \left[
    \int_{\nu_{o}}^{\infty} d\nu (1-e^{\tau_{\rm tot}})-(1-e^{\tau_{\rm
    IGM}}) \right] 
\end{equation}
which measures the 21cm absorption due to the mini--halo in excess of
that due to the IGM. It falls rapidly with $\hat{K}$ and is extremely
small for the most likely value of $\hat{K}=50$. Figures
\ref{fig:plot_21cm_profile} and \ref{fig:EW_21cm} can be compared with
Fig. 1 and 2 of \scite{furlanetto} (although note that in
Fig. \ref{fig:plot_21cm_profile}, we plot $\tau$ as a function of
observed rather than intrinsic frequency).  An entropy floor greatly
reduces the cross-sectional area over which an observable absorption
signal may be detected. If one detects significant mini--halo
absorption over a large comoving patch along the line of sight to a
radio source, one can place an upper limit on the entropy floor
there. This in turn places an upper limit on the high-redshift X-ray
background (from equation~\ref{eqn:KIGM_Xray}), as well as a lower
limit on the redshift at which that patch was first reionized (since
fossils from higher redshift have lower entropy, as in Fig
\ref{fig:entropy_params}).

The chief driver of high-redshift 21cm proposals has of course been to
observe the IGM itself in 21cm emission
\cite{MMR,tozzietal,ciardi_madau}. We note that if the filling factor
of fossil HII regions is large, conditions are very favorable for such
observations. Provided $T_{S} \rightarrow T_{K}$ in the IGM, the only
condition for the IGM to be seen in 21cm emission against the CMB is
for $T_{K} \gg T_{CMB}(z_{\rm obs})$. This is certainly satisfied in
fossil HII regions. It was previously thought that recoil heating from
Ly$\alpha$ photons could heat neutral regions \cite{MMR}, but a
detailed calculation \cite{chen_jordi} shows the heating to be
insufficient. In that case, a high-redshift X-ray background would be
required to heat neutral regions. Such concerns are moot if a period
of early reionization took place (as seems to be indicated by {\it
WMAP} observations), followed by recombination: large tracts of warm,
largely neutral gas would exist. Note, however, that brightness
temperature fluctuations can only be detected if the contribution from
unresolved radio point sources \cite{tozzietal,oh_mack} can be
successfully removed by using spectral structure in frequency space.

\section{Effects on Global Reionization Scenarios}
\label{sec:reionhist}
\subsection{General Considerations}
\label{subsec:reion-general}

We next address the effect of the suppression of mini--halo formation
on the global reionization history.  The importance of the effect
depends on (1) the synchronization of the formation of mini--halos
$t_{\rm sync}$ relative to the typical lifetime $t_{\rm ion}$ of the
ionizing source, which determines the fraction of mini--halos
subjected to feedback, as well as (2) on the recombination
time $t_{\rm rec}$ relative to $t_{\rm sync}$, which determines how
long the ionized regions last and thus the fraction of active (as
opposed to fossil) ionized regions at any given time.

In the limit of $t_{\rm sync}\ll t_{\rm ion}$, the fossil HII regions
would appear only after all the mini--halos had already formed, and
hence the feedback would have no effect on the total amount of
reionization by mini--halos.  In the absence of any other feedback
effects, the IGM could then, in principle, be fully reionized by
mini--halos, given a high enough ionizing photon production efficiency
(although would subsequently recombine).

In the opposite limit, $t_{\rm sync}\gg t_{\rm ion}$, the contribution
of mini--halos to reionization will be strongly suppressed.  We can
obtain a rough estimate for the maximum fraction of the IGM that can
be ionized by the mini--halos.  During the lifetime of its resident
ionizing source, each mini--halo produces an ionized volume $V_{\rm
HII}$, which will correspond to the total number $N_\gamma$ of
ionizing photons injected into the IGM, i.e. $\bar n V\approx
N_\gamma$, where $\bar n$ is the mean hydrogen density.  To a good
approximation, recombinations can be ignored in this phase, since
massive stars have lifetimes shorter than the recombination time,
$t_{\rm rec}\approx 3\times 10^7 [C(1+z)/25] ^{-3}$ years (see
Fig.~\ref{fig:reion_rad} below). On the other hand, the recombination
time is shorter than the Hubble time, $t_{\rm rec}/t_{\rm hub}\approx
[C(1+z)/11] ^{-1.5}$, so in general, at high redshift $z>10$, and for
clumping factors $C>1$, the ionized volume will recombine once its
driving source turns off.

As argued above, for the purpose of the mini--halo suppression, we may
nevertheless simply imagine that the HII region never recombines.
Ignoring recombinations inside the ionized region, this is equivalent
to setting the size of the fossil HII region to be the maximum
(comoving) size of the active HII region, reached at the time when the
ionizing source turns off.

The evolutions of the radii of the active and fossil HII regions are
illustrated in Figure~\ref{fig:reion_rad} for a single $10^3~{\rm
M_\odot}$ metal--free star that turns on at $z=25$.  We follow the
expansion of the ionization front $R_{\rm S}$ by solving the standard
differential equation, taking into account ionizations,
recombinations, and the Hubble expansion (e.g. Shapiro \& Giroux 1987;
Cen \& Haiman 2000).  We assume a constant clumping factor of $C=1$
(upper solid curve) or $C=10$ (lower solid curve). We assume further
that the metal free star emits $40,000$ ionizing photons per stellar
proton at a constant rate for $\sim 3\times10^6$ years (Tumlinson \&
Shull 2000; Bromm, Kudritzki \& Loeb 2001; Schaerer 2002).  Once the
ionizing source turns off, the solid curves show the formal solution
for the evolution of the ionization front. \footnote{This formal
solution predicts a shrinking of the ionization front.  In reality,
once the source turns off, recombinations throughout the ionized
region will decrease the ionized fraction uniformly by the factor
$[R_{\rm S}/R_{\rm S, max}]^3$, rather than reduce the size of a
highly ionized volume.  However, for our purposes of computing the
global ionized fraction, the two interpretations are equivalent.}  The
dashed curves show the size of the fossil HII region.

We can next consider the evolution of the volume filling factor of the
fossil HII regions from the ensemble of mini--halos, and define the
epoch $z_{\rm f}$ when the filling factor of these fossils reaches
unity.  As argued above, no new mini--halos can form at $z<z_{\rm f}$.
At this epoch, the global ionized fraction will be smaller than unity,
because each fossil HII region has already partially recombined.  Thus the
global ionized fraction will be $\sim \exp (-t_{\rm sync}/\bar t_{\rm
rec})$, assuming that the sources typically turned off a time $t_{\rm
sync}$ ago, and $\bar t_{\rm rec}$ is the average recombination time
over the interval $t_{\rm sync}$.  For example, assuming that the
fossil HII regions overlap at $z_f=20$, with the typical sources born
at $z=25$, and the recombination time evaluated at $z=22.5$ (with
$C=1$), we find $t_{\rm sync}\approx \bar t_{\rm rec}\approx 5\times
10^7$ years, and a maximal ionized fraction of $\sim 30\%$.

\begin{figure}
\psfig{file=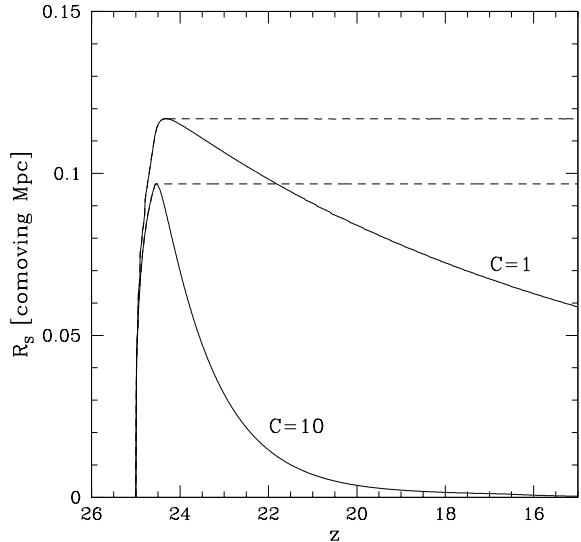,width=80mm}
\caption{The evolution of comoving radius of the ionized region around
a single $10^3~{\rm M_\odot}$ metal--free star that turns on at
$z=25$, and is assumed to emit $40,000$ ionizing photons in $\sim
3\times10^6$ years. The upper and lower solid curves show solutions
that include recombinations with $C=1$ and $C=10$. The ionized volume
shrinks (or equivalently, the ionized fraction is decreased by the
factor $[R_{\rm S}/R_{\rm S, max}]^3$) after the source turns off at
$z\sim 24$.  The corresponding dashed curves show the radii of the
fossil HII regions; these are assumed to stall at the maximum radius
of the HII sphere and stay constant thereafter.}
\label{fig:reion_rad}
\end{figure} 

\subsection{Models for the Reionization History}
\label{subsec:reion-models}

To refine the above considerations, we next compute the evolution of
the global ionized fraction using a semi--analytical model adopted
from Haiman \& Holder (2003, hereafter HH03).  For technical details,
the reader is referred to that paper.  Here we only briefly summarize
the main ingredients of the model, and describe the modifications we
made to include the suppression of the formation of new mini--halos
inside fossil HII regions.

In this model, we follow the volume filling fraction $F_{\rm HII}$ of
ionized regions, assuming that discrete ionized Str\"omgren spheres
are being driven into the IGM by ionizing sources located in dark
matter halos.  For simplicity, we ignore Helium in this work.  The
dark matter halo mass function is adopted from Jenkins et
al. (2001). We consider two distinct types of ionizing sources,
located in halos with virial temperatures of $10^2\,{\rm K} \lsim
T_{\rm vir} \lsim 10^4$K (mini--halos) or $T_{\rm vir} \gsim 10^4$K
(large halos).  We assume that in mini--halos, a fraction $f_*=0.005$
of the baryons turn into metal free stars (Abel, Bryan \& Norman 2000,
2002; Bromm, Coppi \& Larson 1999, 2002), which produce
$N_\gamma=40,000$ ionizing photons per baryon, and $f_{\rm esc}=100\%$
of these photons escape into the IGM.  The product $\epsilon\equiv f_*
f_{\rm esc} N_\gamma=200$ determines the size of the ionized regions
around mini--halos.  For large halos, we adopt $f_*=0.1$, $f_{\rm
esc}=20\%$, and $N_\gamma=4,000$, for a total efficiency
$\epsilon=80$.  In contrast with HH03, we here do not distinguish
halos with $10^4~{\rm K}<T_{\rm vir}<2\times10^5~{\rm K}$ from $T_{\rm
vir}>2\times10^5~{\rm K}$ (see Dijkstra et al. 2003 for the reason why
this distinction is likely unimportant).

\begin{figure}
\psfig{file=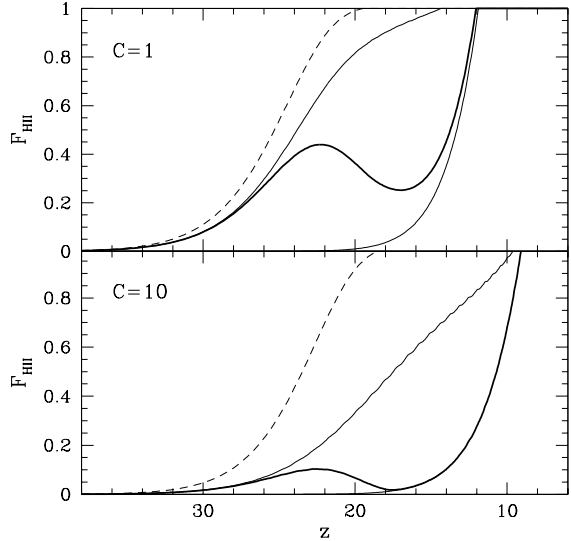,width=80mm}
\caption{The evolution of the volume filling fraction $F_{\rm HII}$ of
ionized hydrogen in reionization models with different assumptions
about feedback on mini--halo formation. The upper and lower panels
assume constant clumping factors of $C=1$ and $C=10$, respectively.
In both panels, the dashed curves show the volume filling fraction of
the fossil HII regions.  The thick solid curves show the evolution of
$F_{\rm HII}$ assuming that mini--halos cannot form inside fossil HII
regions.  The upper solid curves assume that mini--halos are only
excluded from forming inside active ionized regions. Finally, the
lower solid curves ignore the contribution to $F_{\rm HII}$ from any
mini--halos.}
\label{fig:reion_xe}
\end{figure}

As in HH03, we then follow the evolution of the global ionized
fraction $F_{\rm HII}(z)$ by summing the ionized volumes surrounding
the individual dark matter halos. We exclude the formation of
mini--halos in active ionized regions (but assume larger halos are
impervious to photoionization feedback).  The evolution of $F_{\rm
HII}$ in this model is shown by the upper thin solid curve in
Figure~\ref{fig:reion_xe} (with $C=1$ in the upper panel, and $C=10$
in the lower panel).  As the figure shows, in the absence of other
feedback effects (see discussion in Haiman, Abel \& Rees 2000 and
Haiman 2003 for other feedback effects), the mini--halos could ionize
the IGM in full by $z\sim 14$ ($C=1$), or nearly fully by $z\sim 10$
(if clumping is assumed to be more significant, $C=10$).

We next compute $F_{\rm HII}$ in the same model, except that we
exclude the formation of new mini--halos in fossil HII regions.  This
is easily accomplished in practice: the suppression factor ($1-F_{\rm
HII}$) for the formation rate of mini--halos is replaced by a factor
($1-F^\prime_{\rm HII}$). Here the fossil filling factor
$F^\prime_{\rm HII}$ is computed the same way as the ionized fraction
$F_{\rm HII}$, except that the individual ionized regions are assumed
to follow the dashed curves from Figure~\ref{fig:reion_rad} rather
than the solid curves.  In Figure~\ref{fig:reion_xe}, we show the
evolution of the volume filling factor $F^\prime_{\rm HII}$ of the
fossil HII regions (dashed curves), as well as the reionization
history $F_{\rm HII}(z)$ as the thick solid curves.  Finally, for
reference, we show $F_{\rm HII}(z)$ with mini--halos completely
excluded ($\epsilon=0$) as the lower thin solid curves.

As apparent from Figure~\ref{fig:reion_rad}, the exclusion of new
mini--halo formation from the fossil HII regions causes a significant
suppression of the total ionized fraction that can be reached by
mini--halos.  Under the rather optimistic set of assumptions described
by the thick solid curve in the upper panel, the maximum ionized
fraction that can be reached is $\sim 40\%$.  In reality, clumping is
unlikely to be unity in the immediate vicinity ($\lsim 100$ kpc) of
the ionizing sources (Haiman, Abel \& Madau 2000).

The total electron scattering optical depth attributable to
mini--halos (the appropriately weighted integral between the thick
solid curve and the lower thin solid curve) is $\tau=0.07$ and
$\tau=0.014$ in the $C=1$ and $C=10$ cases, respectively.  This makes
it unlikely that mini--halos can fully account for the large optical
depth $\tau=0.17$ measured by {\it WMAP}. Note the thick curve in the
upper panel of Figure~\ref{fig:reion_xe} has a total $\tau=0.2$
($\tau=0.07$ attributable to minihalos, and $\tau=0.13$ to larger
halos), so that it is consistent with the {\it WMAP} measurement.
Note that raising the efficiencies in mini--halos would not increase
the optical depth attributed to minihalos, since feedback then would
set in earlier (we have explicitly verified that $\tau$ is
approximately independent of efficiencies over a range of
multiplicative factors 0.1 - 10 for $\epsilon$).  Finally, note that
the suppression considered here and shown in Figure~\ref{fig:reion_xe}
provides a negative feedback {\it in addition} to the negative
feedback expected from ${\rm H_2}$ photodissociation (Haiman, Rees \&
Loeb 1997; Haiman, Abel \& Rees 2000).  In fact, as discussed above,
preheating {\it amplifies} the effect of the ${\rm
H_2}$-photodissociative negative feedback.

Because an entropy floor essentially eliminates gas clumping due to
mini-halos (as shown in section \ref{subsection:coll_clump}), the
reionization history is likely to more closely approximate the C=1
case than the C=10 case until low redshifts $z <10$ (when gas clumping
due to larger structures predominates). An entropy floor therefore has
two countervailing effects on reionization: by suppressing star
formation in mini--halos, it reduces the comoving emissivity. However,
by reducing gas clumping, it also reduces the photon budget required
for reionization. It is interesting to note that that the evolution of
the filling factor is non-monotonic for the C=1 case: feedback due to
early reionization naturally produces a bump in the comoving
emissivity, and a pause ensues before larger halos (which can resist
feedback) collapse. This is similar to the reionization histories
derived by \cite{cen03,wl03}, but is regulated by feedback from the
entropy floor rather than a Pop III to Pop II transition due to a
universal metallicity increase.

\section{Conclusions}
\label{section:conclusions}

In this paper, motivated by the {\it WMAP} results, we have considered
the feedback effect of early reionization/preheating on structure
formation. This feedback effect is {\it inevitable} in any
reionization scenario in which star formation throughout the universe
is not completely synchronized. Our principal conclusions are as
follows:

1. Fossil HII regions have a residual entropy floor after
   recombination and Compton cooling which is {\it higher} than the
   shock entropy for mini--halos ($T_{\rm vir}< 10^{4}$K); thus, such
   halos accrete gas isentropically. The IGM entropy depends primarily
   on the redshift and only weakly on the overdensity $\delta$; it is
   thus largely independent of the details of structure formation. For
   this reason, and also because it is conserved during adiabatic
   accretion or Hubble expansion, the gas entropy is a more
   fundamental variable to track than the temperature. We provide a
   simple analytic formula for the temperature (and hence entropy
   $K=T/n^{2/3}$), in equations \ref{eqn:T_analytic} and
   \ref{eqn:t_rec}. An early X-ray background would also heat the
   entire IGM to similarly high adiabats. The entropy floor due to the
   latter would be much more spatially uniform.

2. We apply the entropy formalism used to calculate the effect of
   preheating on low-redshift galaxy clusters to the high--redshift
   minihalos. We obtain detailed gas density and pressure profiles,
   which we use to calculate the impact of preheating on the central
   density, accreted gas fraction, gas clumping factor, and mini--halo
   baryonic mass function.

3. These quantities can then be used to calculate global effects of
   preheating. The collapsed gas fraction in minihalos falls by $\sim
   2$ orders of magnitude, while the gas clumping factor falls to $C
   \rightarrow 1$, as for a uniform IGM. An entropy floor reduces the
   photon budget required for full reionization by about an order of
   magnitude, by reducing gas clumping and eliminating the need for
   mini--halos to be photo-evaporated before reionization can be
   completed.

4. However, an entropy floor does not necessarily promote early
   reionization: it also sharply reduces the comoving emissivity. By
   reducing the central gas densities in mini--halos, preheating
   impedes $\htwo$ formation and cooling, and reduces the critical UV
   background required for $\htwo$ suppression by 2-4 orders of
   magnitude. {\em Thus, once a comoving patch of the IGM is
   reionized, no subsequent star formation in mini--halos can take
   place in that volume.} The patch can only be reionized by more
   massive halos $T>10^{4}$K, which can undergo atomic cooling. By
   furnishing an entropy floor, X-rays also suppress $\htwo$
   formation. Thus, contrary to conventional wisdom, X-rays provide
   negative rather than positive feedback for early star formation.

5. Mini-halos will not be observable in 21cm emission/absorption in
   fossil HII regions. Thus, 21cm observations provide an unusual
   probe of the topology of reionization: mini-halos trace out regions
   of the IGM which have {\it never} been ionized. If mini-halos are seen in
   large numbers, this places an upper limit on the filling factor of
   fossil HII regions and the X-ray background at that redshift. 

6. We have computed the reionization histories as in HH03, but taking
   the feedback effect of an entropy floor (and reduction of gas
   clumping) into account. The strong feedback in fossil HII regions
   imply that HII fronts at high redshift never overlap, and global
   reionization at high redshift does not occur. This limits the contribution of   mini--halos to the reionization optical depth $\tau < 0.07$, almost
   independent of star formation efficiency in mini--halos (if star
   formation is more efficient, feedback sets in earlier). Thus, the
   bulk of the optical depth observed by WMAP must come from more
   massive objects. 

 7. Strikingly, we obtain a double-peaked reionization history: an
   early peak in which the universe is filled with fossil HII regions,
   followed by a pause before more massive halos collapse which
   finally fully reionize the universe. This is similar to 'double
   reionization' scenarios computed by other authors
   (e.g. \scite{cen03,wl03}), but one in which the the comoving
   emissivity is regulated by gas entropy, rather than a Pop III to
   Pop II transition due to a universal metallicity increase.  This
   last point deserves additional comment. A metallicity-regulated
   evolution of the emissivity requires that metal pollution is fairly
   spatially uniform. However, different parts of the IGM likely
   undergo the Pop III to Pop II transition at different
   epochs. Furthermore, an increase in metallicity does not
   necessarily result in a drop in the overall emissivity: metal line
   cooling likely results in a larger star formation efficiency in
   halos, since metals are not subject to internal UV
   photo-dissociation, unlike $\htwo$. The factor of $\sim$10 drop in
   the HI ionizing emissivity per stellar baryon could be outweighed
   by the increase in the total mass of stars formed.  In comparison,
   we argue that an entropy-regulated transition is inevitable, and
   therefore more robust.  This is an important conclusion, given the
   possibility that future CMB polarization studies will be able to
   distinguish among different reionization histories (HH03; Holder et
   al 2003).

In this semi-analytic study, we essentially assumed a single value of
the entropy floor at each redshift, for all halos. In reality, as we
discussed above, there should be large fluctuations in entropy,
depending on the topology/history of reionization and the
accretion/merger history of halos. It would be interesting to study
these effects in detail in three dimensional numerical simulations. To
begin with, it would be interesting to re-run existing simulations of
the collapse of gas in mini--halos, but adding an entropy floor in the
initial conditions. Such studies could quantify more precisely the
effects of an entropy floor in suppressing $\htwo$ formation and
cooling in mini--halos. In fact, since densities are lower and gas
cooling is reduced, it should be a computationally more tractable
problem. With the aid of larger volumes, several global issues
mentioned above could be addressed: for instance, a global,
self-consistent study of feedback as in \scite{machacek}, but
including the effects of both the entropy floor {\it and} UV
feedback. Because of its strong effect on the evolution of the
comoving emissivity, gas entropy acts as a self-regulating mechanism
which likely has a strong influence in controlling the progress of
reionization.

\section*{Appendix I: Analytic expression for the final temperature}

It is useful to have an approximate analytic expression for the final
temperature a parcel of gas cools down to, given the initial
temperature $T_{i}$, redshift $z$, overdensity $\delta$ and length of
time spent cooling $t$. This allows one to
quickly estimate the effect of early reionization in different
situations without evolving the full chemistry code. We develop such
an expression in this Appendix. 

At the redshifts and overdensities of interest, Compton cooling
dominates by far. The most relevant physics is Compton cooling and
hydrogen recombination:
\begin{eqnarray}
\dot{T}&&=-\frac{8}{3} \frac{a T_{\gamma}^{4} \sigma_{T}}{m_{e} c}
\frac{x_{e}}{1+x_{e}}(T-T_{\gamma}) \\ \nonumber
\dot{x_{e}}&&=-x_{e}^{2} n \alpha
\end{eqnarray}
Note that $(T-T_{\gamma}) \approx T$ since $T \gg T_{\gamma}$, and
$\alpha \propto T^{-0.7}$ in the temperature range of
interest. Assuming the gas is fully ionized $x_{e}=1$ at some initial
temperature $T_{i}$, we obtain the analytic solution:
\begin{equation}
T(x_{e})=T_{i}\left[1+1.4 A \left( {\rm ln}\left(\frac{1+x_{e}}{x_{e}}
  \right) - {\rm ln}(2) \right) \right]^{-1/0.7} 
\label{eqn:T_analytic}
\end{equation} 
where
\begin{equation}
A \equiv \frac{4}{3} \frac{a T_{\gamma}^{4} \sigma_{T}}{n
  \alpha(T_{i}) m_{e} c} = \frac{t_{rec}(T_{i})}{t_{C}} \propto
  \frac{(1+z)}{\delta} T_{i}^{0.7} 
\end{equation}
This gives temperature as a function of ionization fraction
$x_{e}$. We therefore need to know the final ionization fraction. How
can we estimate it? If the gas recombines isothermally at temperature
$T^{\prime}$, the ionization fraction is given by:
\begin{equation}
x_{e}(t)=\frac{x_{o}}{1+(t/t_{rec}(T^{\prime}))}
\label{eqn:t_rec}
\end{equation}
where $x_{o}=1$ is the initial ionization fraction. If the gas cools
as it recombines, substituting the instantaneous temperature
$T^{\prime}$ into the expression, would overestimate the speed of
recombination and underestimate $x_{e}$. We find that if we substitute
$T^{\prime}=2T$ and substitute this into equation
\ref{eqn:T_analytic}, the solution of this non-linear equation for
$T_{f}$ is remarkably close to the full non-equilibrium solution (see
points in Figure \ref{fig:temp_params}). We have also verified that we
obtain fairly accurate results for $x_{e}(t)$. Equations
\ref{eqn:T_analytic} and \ref{eqn:t_rec} thus give
$T_{f}(z,\delta,T_{i},t)$.

Our neglect of recombination line cooling fails in high density
regions. This leads to at most a factor $\sim 2$ error in the final
temperature, since recombination line cooling can cool gas down to at
most $\sim 5000$K. The entire discussion assumes that ${\rm H_{2}}$
formation and cooling is not competitive with Compton cooling, which
is generally true in low density regions.


\begin{thebibliography}{} 
 
\bibitem[Abel, Bryan \& Norman <2000>]{abel} Abel T., Bryan
     G.L., Norman M.L., 2000, ApJ, 540, 39

\bibitem[Abel, Bryan \& Norman <2002>]{abn02} Abel, T., Bryan, G. L.,
  \& Norman, M. L. 2002, Science, 295, 93

\bibitem[Allison \& Dalgarno <1969>]{allison} Allison, A.C., \&
  Dalgarno, A., 1969, ApJ, 158, 423

\bibitem[Babul et al <2002>]{babuletal} Babul, A., Balogh, M.L.,
Lewis, G.F., Poole, G.B., 2002, MNRAS, 330, 329

\bibitem[Barkana \& Loeb <1999>]{barkanaloeb99} Barkana, R., \& Loeb,
  A., 1999, ApJ, 523, 54

\bibitem[Barkana \& Loeb <2002>]{barkana_loeb_review} Barkana, R., \& Loeb,
  A., 2002, ApJ, 578, 1

\bibitem[Barkana \& Loeb <2002>]{barkanaloeb} Barkana, R., \& Loeb,
  A., 2002, ApJ, 578, 1

\bibitem[Balogh, Babul \& Patton <1999>]{baloghetal} Balogh, M.L.,
Babul, A., \& Patton, D.R., 1999, MNRAS, 307, 463

\bibitem[Benson et al <2002>]{benson} Benson, A.J., Lacey, C.G.,
Baugh, C.M., Cole, S., Frenk, C.S., 2002, MNRAS, 333, 156

\bibitem[Bromm, Coppi, \& Larson <1999>]{bcl01} Bromm, V., Coppi,
P. S., \& Larson, R. B. 1999, ApJ, 527, 5 

\bibitem[Bromm et al <2001>]{bromm01}  Bromm, V., Ferrara, A., Coppi,
P.S., \& Larson, R.B. 2001, MNRAS, 328, 969

\bibitem[Bromm, Coppi \& Larson <2002>]{bromm} Bromm V., Coppi P.S.,
     Larson R.B., 2002, ApJ, 564, 23 

\bibitem[Bromm, Kudritzki, \& Loeb <2001>]{bkl01} Bromm, V.,
Kudritzki, R. P., \& Loeb, A. 2001, ApJ, 552, 464

\bibitem[Bullock, Kravtsov \& Weinberg <2000>]{bullock} Bullock, J.S.,
Kravtsov, A.V., \& Weinberg, D.H., 2000, ApJ, 539, 517

\bibitem[Cen <2003>]{cen03} Cen, R. 2003, ApJ, submitted, astro-ph/0210473 

\bibitem[Cen \& Haiman <2000>]{ch00} Cen, R., \& Haiman, Z. 2000, ApJ, 542, L75

\bibitem[Chen \& Miralda-Escude <2003>]{chen_jordi} Chen, X., \&
  Miralda-Escude, J., 2003, ApJ, submiteed, astro-ph/0303395

\bibitem[Ciardi, Ferrara \& Abel <2000>]{ciardi} Ciardi, B., Ferrara,
A., \& Abel, T. 2000, ApJ, 533, 594

\bibitem[Ciardi \& Madau <2003>]{ciardi_madau} Ciardi, B., \& Madau,
P., 2003, ApJ, submitted, astro-ph/0303249

\bibitem[Dijkstra et al <2003>]{dhr03} Dijkstra, M., et al. 2003, in preparation

\bibitem[Elvis et al <1994>]{elvis94} Elvis, M., et al, 1994, ApJS,
95, 1

\bibitem[Gnedin <2000>]{gnedin00} Gnedin, N.Y., 2000, ApJ, 542, 535

\bibitem[Fixsen \& Mather <2002>]{fixmath} Fixsen, D. J., \& Mather, J. C. 2002, ApJ, 581, 817

\bibitem[Frenk et al <1999>]{frenketal} Frenk, C., et al, 1999, ApJ,
  525, 554

\bibitem[Furlanetto \& Loeb <2002>]{furlanetto} Furlanetto, S.R., \&
  Loeb, A., 2002, ApJ, 571, 1

\bibitem[Haiman <2003>]{hz03} Haiman, Z. 2003, in Carnegie Symposium Review Series I, ed. L. Ho.

\bibitem[Haiman, Abel \& Rees <2000>]{haimanh2} Haiman, Z., Abel, T., \& Rees, M. J. 2000, ApJ, 534, 11

\bibitem[Haiman, Abel \& Madau <2001>]{haimanabelmadau} Haiman, Z.,
Abel, T., \& Madau, P. 2001, ApJ, 551, 559

\bibitem[Haiman \& Holder <2003>]{hh03} Haiman, Z., \& Holder,
G. 2003, ApJ, in press, astro--ph/0302403 (HH03)

\bibitem[Haiman, Rees \& Loeb <1996>]{hrl96} Haiman, Z., Rees, M. J., \& Loeb, A. 1996, ApJ, 467, 522

\bibitem[Haiman, Rees \& Loeb <1997>]{hrl97} Haiman, Z., Rees, M. J., \& Loeb, A. 1997, ApJ, 476, 458 [erratum: 1997, ApJ, 484, 985]

\bibitem[Haiman, Spergel \& Turner <2003>]{haiman_midIR} Haiman, Z.,
  Spergel, D.N., \& Turner, E.L., 2003, ApJ, 585, 630

\bibitem[Hansen \& Haiman <2003>]{hansen} Hansen, S., \& Haiman,
Z. 2003, ApJL, submitted, astro-ph/0305126

\bibitem[Hellsten \& Lin <1997>]{hellsten} Hellsten, U., \& Lin,
D.N.C., 1997, astro--ph/9708086 

\bibitem[Holder et al <2003>]{holder03} Holder, G., Haiman, Z., Kaplinghat, M., \& Knox, L. 2003,  ApJ, in press, astro--ph/0302403 

\bibitem[Iliev et al <2002a>]{mh1} Iliev, I.T., Shapiro, P.R.,
  Ferrara, A., \& Martel, H., 2002a, ApJ, 572, L123

\bibitem[Iliev et al <2002b>]{mh2} Iliev, I.T., Scannapieco, E.,
  Martel, H., \& Shapiro, P.R., 2002b, MNRAS, submitted, astro-ph/0209216

\bibitem[Kogut et al <2003>]{kogut} Kogut A. et al, 2003, ApJ,
     submitted; astro-ph/0302213

\bibitem[Knox et al <1999>]{knoxetal99} Knox, L., Scoccimarro, R., \&
  Dodelson, S., 1998, PhRvL, 81, 2004

\bibitem[Machacek, Bryan \& Abel <2001>]{machacek} Machacek, M. E.,
  Bryan, G. L., Abel, T. 2001, ApJ, 548, 509

\bibitem[Madau, Meiskin \& Rees <1997>]{MMR} Madau, P., Meiskin A., \&
Rees, M.J., 1997, ApJ, 475, 429  

\bibitem[Madau, Ferrara \& Rees <2001>]{madau01} Madau, P., Ferrara,
A., \& Rees, M.J. 2001, ApJ, 554, 92

\bibitem[Magliocchetti, Salvaterra \& Ferrara <2003>]{ferrara}
Magliocchetti, M., Salvaterra,R., \& Ferrara, A., 2003, MNRAS, 342, L25 

\bibitem[Navarro, Frenk \& White <1997>]{NFW} Navarro, J.F., Frenk,
C.S., \& White, S.D.M., 1997, ApJ, 490, 493

\bibitem[Oh <1999>]{oh99} Oh, S.P., 1999, ApJ, 527, 16

\bibitem[Oh <2000>]{oh2000} Oh, S.P., 2000, PhD thesis, Princeton
University, Chapter 4.

\bibitem[Oh <2001>]{oh2001} Oh, S.P., 2001, ApJ, 553, 25

\bibitem[Oh \& Benson <2003>]{oh_benson} Oh, S.P., \& Benson, A.,
  2003, MNRAS, 342, 664

\bibitem[Oh \& Haiman <2002>]{oh_haiman} Oh, S.P., \& Haiman, Z., 2002,
  ApJ, 569, 558 (OH02)

\bibitem[Oh \& Mack <2003>]{oh_mack} Oh, S.P., \& Mack, K., 2003,
  MNRAS, submitted, astro-ph/0302099

\bibitem[Oh, Cooray \& Kamionkowski <2003>]{ohetal03} Oh, S.P., Cooray,
  A., \& Kamionkowski, M., 2003, MNRAS, 342, L20

\bibitem[Ricotti, Gnedin \& Shull <2002>]{ricotti} Ricotti, M.,
Gnedin, N.Y., \& Shull, J.M., 2002, 575, 49

\bibitem[Rosati, Borgani \& Norman <2002>]{rosatietal2002} Rosati, P.,
Borgani, S., \& Norman, C., 2002, ARAA

\bibitem[Santos, Bromm \& Kamionkowski <2002>]{santos_mike} Santos,
M.R., Bromm, V., Kamionkowski, M., 2002, MNRAS, 336, 1082

\bibitem[Santos <2003>]{santos}  Santos, M. G., Cooray, A., Haiman, Z., Knox, L., \&  Ma, C.-P. 2003, ApJ, submitted, astro-ph/0305471

\bibitem[Schaerer <2002>]{sc02} Schaerer, D. 2002, A\&A, 382, 28

\bibitem[Shapiro \& Giroux <1987>]{sg87} Shapiro, P., \& Giroux, M. L. 1987, ApJ, 321, L107

\bibitem[Shapiro, Iliev \& Raga <1999>]{shapiro99} Shapiro, P.R., Iliev, I.T., Raga, A.C., 1999, MNRAS, 307, 203 

\bibitem[Shapiro, Raga \& Mellema <1998>]{shapiro98} Shapiro, P. R.,
Raga, A. C., \& Mellema, G. 1998, in Molecular Hydrogen in the Early
Universe, Memorie Della Societa Astronomica Italiana, Vol. 69,
ed. E. Corbelli, D. Galli, and F. Palla (Florence:
Soc. Ast. Italiana), p. 463

\bibitem[Shapiro et al <2003>]{shapiro03} Shapiro, P.R., Iliev, I.T.,
Raga, A.C., Martel, H., 2003, in The Emergence of Cosmic Structure, 13th
Annual October Astrophysics Conference in Maryland, S. Holt \&
C. Reynolds, eds (AIP), in press, astro-ph/0302339 

\bibitem[Shull \& van Steenberg <1985>]{shull} Shull, J.M., \& van Steenberg, M.E., 1985, ApJ, 298, 268

\bibitem[Somerville <2002>]{somerville} Somerville, R.S., 2002, ApJ,
572, L23 

\bibitem[Spergel et al <2003>]{spergel} Spergel D.N. et al, 2003,
     ApJ, submitted, astro-ph/0302209

\bibitem[Tegmark et al <1997>]{tegmark} Tegmark, M., Silk, J., Rees, M.J., Blanchard, A., Abel, T., Palla, F., 1997, ApJ, 474, 1 

\bibitem[Tozzi \& Norman <2001>]{tozzinorman} Tozzi, P., \& Norman,
  C., 2001, ApJ, 546, 63

\bibitem[Tozzi et al <2000>]{tozzietal} Tozzi, P., Madau, P., Meiskin,
A., \& Rees, M.J., 2000, ApJ, 528, 597

\bibitem[Tumlinson \& Shull <2000>]{ts00} Tumlinson, J., \& Shull,
M. J. 2000, ApJ, 528, L65

\bibitem[Venketesan et al <2001>]{venkatesan} Venkatesan, A., Giroux, M.L., Shull,
J.M., 2001, ApJ, 563, 1

\bibitem[Voit et al <2002>]{voitetal} Voit, G.M., Bryan, G.L., Balogh,
  M.L., Bower, R.G., 2002, ApJ, in press, astro-ph/0205240

\bibitem[Wyithe \& Loeb <2003>]{wl03} Wyithe, S., \& Loeb, A. 2003, ApJ, 586, 693

\end{thebibliography}
\end{document}